\documentstyle[preprint,prd,aps,eqsecnum]{revtex}
\begin{document}
\draft
\title{Gravitational waves from inspiralling compact binaries:
Angular momentum flux, evolution of the orbital 
elements and the wave form to the second
post-Newtonian order}
\author{A. Gopakumar and Bala R. Iyer}
\address{ Raman Research Institute, Bangalore 560 080, India}
\date{\today}
\maketitle
\begin{abstract}
The post-post-Newtonian (2PN)
accurate mass quadrupole moment, for
compact binaries of arbitrary mass ratio, moving in general
orbits is obtained  by the multi-polar post Minkowskian 
approach of Blanchet, Damour, and Iyer~(BDI).
Using this, for binaries in general orbits, 
 the 2PN contributions  to the gravitational waveform,
 and the associated  far-zone energy and angular momentum fluxes 
are computed. For quasi-elliptic orbits, 
the energy and angular momentum fluxes  are averaged over
an orbital period, and employed to determine
 the 2PN corrections to the rate of decay of the orbital
elements. 
\end{abstract}
\widetext
\baselineskip22pt
\section{Introduction}
Inspiralling compact binaries are one of 
the most promising sources of 
gravitational radiation for kilometer size laser interferometric
gravitational wave detectors
like  LIGO\cite{ligo} and VIRGO\cite{virgo}.
The method of matched filtering will be employed to
detect and extract information of binaries from the inspiral waveforms
\cite{KT87,BFS91}. In
this technique one cross correlates the noisy output of a detector
with
theoretical templates. For this technique to be successful, the
templates must
remain in phase with the exact -- general relativistic --  waveform
as long as
possible. If the signal and template lose phase
with each other even by  a
cycle in the ten  thousand as the waves sweep through the bandwidth
of the detector  their cross-correlation will be significantly
reduced
and one  may lose the event altogether.
Detailed works on data analysis aspects
\cite{curtprl,finncher93,curtena,bsd} bears out this inference
and one is forced to a  description of the evolution of the
binary system,  using the best available theory of gravity
to substantially higher
accuracy than that provided by the lowest order
Newtonian approximation.
The  construction  of
accurate theoretical templates for  inspiralling compact binaries
involves the solution to two different but related aspects
referred to respectively as the ``wave generation problem''
and the ``radiation reaction problem''.
In the generation problem one computes the gravitational
waveforms and the associated energy and angular momentum fluxes
emitted by the binary for a  fixed, specified  orbital
motion ignoring the back reaction of the radiation emission
on the orbit.  In the radiation reaction problem on the
other hand, one computes the
effect of the emitted radiation on  the orbital phase evolution
and this is of crucial importance  as explained earlier.

Einstein's\cite{AE1916} far field quadrupole 
equation is the solution to
the generation problem to the lowest order but applies only
to objects held together by non-gravitational forces.
 Fock\cite{FOCKT} and Landau-Lifshiftz\cite{LLT} 
provided two very different
methods to generalize the above results to weakly
self-gravitating systems and the  two approaches are
the starting points for the two methods available today
to calculate gravitational wave generation to higher
orders: the Blanchet Damour Iyer(BDI)\cite{bdigen} approach and the
Epstein Wagoner Thorne Will Wiseman(EWTWW)\cite{wwgen} approach.

The BDI approach builds on a Fock type derivation
using the double-expansion method of Bonnor \cite{WBB59}.
This approach makes a clean separation of the near-zone
and the wave-zone effects. It is mathematically well-defined,
algorithmic and provides corrections to the
quadrupolar formalism in the form of compact support
integrals or more generally well-defined analytically
continued integrals. The scheme has a modular structure :
the  final results are obtained by combining an `external
zone module' with a `near zone module' and a `radiative zone
module'.
For dealing with strongly 
self-gravitating material sources like
neutron stars or black holes one needs to use a `compact body
module' supplemented by an `equation of motion module' to
describe their `conservative' orbital motion.
Using this approach  the generation of gravitational waves
from inspiralling compact binaries of arbitrary mass
ratio moving in a {\em quasi circular} orbit has been computed to
2PN accuracy
\cite{LB95,BDI95} and more recently to 2.5PN accuracy\cite{LB96}.
In this paper, in the first instance,
using the BDI approach, we extend the above 2PN treatment to
inspiralling compact binaries moving in a {\em general orbit}
and compute the 2PN contributions to the waveform and the energy flux.
Unlike for circular orbits,  the  angular momentum flux from general
orbits provides additional, independent information and we 
also compute the same.

                      The Epstein-Wagoner-Thorne-Will-Wiseman 
approach  on the other hand,
builds on a Landau-Lifshiftz type treatment to derive post-
Newtonian corrections to the lowest order quadrupole formula.
The combined use of an effective stress energy tensor for the
gravitational
field ( with non-compact support ) and of formal post-Newtonian
expansions led
to the appearance of divergent integrals.
The presense of the divergent 
integrals and the lack of a clear separation between the near zone
and the wave zone were unsatisfactory features of this scheme until
recently. However, last year, Will and Wiseman\cite{WW96} have provided a
resolution
to this problem by taking literally the statement that the solution
is a {\em retarded integral i.e.,} an integral over the entire past
null cone of the field point. A careful evaluation of the far-zone
contributions, then shows that all integrations are indeed
convergent and finite and moreover the tail terms are also correctly
recovered. Using this treatment, Will and Wiseman have computed
the 2PN accurate waveform and energy flux for {\em general orbits}.
We thus have two approaches to the 2PN generation,
which can provide a useful check on the long and tedious
algebra.

         The most accurate results to date for the generation
and the radiation reaction have been obtained in the  limit   
where a test body orbits a very massive central body.  
In this complementary approach, 
based on black hole perturbation techniques
there exist  numerical results that are exact in $(v/c)$ and analytical
results accurate to the 5.5PN order, {\em i.e.}
corrections of $O[(v/c)^{11}]$
for a test particle
in a circular orbit around a Schwarzschild black hole
\cite{EP93,CFPS93,TN94,TS94,SSTT95,HT95,TSTS96,TTHTMS},
where $v$ is the orbital velocity of the test particle. 
For a test
particle in a slightly eccentric orbit,
around a spinning black hole expressions
for the energy and the angular momentum
fluxes have also been computed to the 2.5PN order\cite{HT95}.

It is well-known that, 
when  gravitational waves from prototype systems like the
binary pulsar 1913+16 
enter the bandwidth of the terrestial interferometric detectors, 
the eccentricity of these binary systems would have been drastically 
reduced and have become negligible due to the gravitational radiation reaction.
A treatment of such systems is simpler since the quasi-circular
approximation for their orbits is amply adequate and the corresponding
waveforms do not depend on the eccentricity parameter of the orbit.
However there exist scenarios in which  the eccentricity is no
longer negligible and this would require the more general treatment
provided in this paper and   independently by Will and Wiseman\cite{WW96}.
One such possibility has been discussed by Shapiro and Teukolsky\cite{ST85}
in the context of the formation of supermassive black holes.
They consider a cluster of
compact objects --neutron stars 
and black holes-- residing at the center of a galactic nucleus.
 Coulomb scattering and dissipative processes will drive 
such a cluster to a high density, high redshift state. 
Once the central redshift becomes sufficiently large,
relativistic instability sets in, and the core 
undergoes catastrophic collapse to
form a supermassive black hole.
Quinlan and Shapiro \cite{QS87} have shown that
during the final year of
the evolution of such a  cluster, just prior to the 
catastrophic collapse, there can be $ 100 - 10^4 $ 
evolving black hole binaries in eccentric orbits driven by gravitational 
radiation reaction, with masses in the range $ 10 - 100\,M_{\odot}$.
Ground based  interferometric 
detectors will be sensitive to the gravitational radiation from these 
binaries in eccentric orbits 
and such eccentric binaries  may prove to be another possible
class of gravitational wave sources. 
More recently,  Flanagan and Hughes\cite{FH97} suggest that 
intermediate mass
black hole binaries of the kind considered above 
--with total masses in the range $ 50 M_{\odot}
\leq M \leq $ (a~few) $\times 10^3 M_{\odot}$-- may well be
the first sources to be detected  by LIGO and VIRGO. 
The other possibility involves compact objects orbiting 
$10^6$ to $10^7 M_{\odot}$  black holes,
that seem fairly  common in  galactic nuclei.
In this case the compact objects could be   scattered
into very eccentric orbits  orbits via gravitational deflections 
by other stars. However,  by the time gravitational radiation reaction becomes
the dominant orbital driving force, 
there is not enough inspiral remaining to fully circularise
these orbits. 
 Hils and Bender\cite{HB95} have argued that the 
event rates for the above process are very encouraging 
and the chances of such signals 
being observed by 
Laser Interferometric Space Antenna, LISA\cite{lisa}  
appear very good.
                          
The expressions for the far-zone 
energy and angular momentum fluxes find application in  another 
related but distinct problem; the  evolution  of the orbital
elements of systems like the binary pulsar 1913+16; most importantly  
the orbital period, and to a lesser extent
the eccentricity and semi-major axis. 
This application, in addition to the generation results discussed
earlier, requires a convenient representation of the post-Newtonian motion
of two point masses in elliptical orbits. To 1PN accuracy, 
such a quasi-Keplerian
representation  has been provided by Damour and Deruelle\cite{DD85},
while to the 2PN order a generalised quasi-Keplerian representation
has been implemented by Damour, Sch\" afer, and Wex\cite{DS87,DS88,SW93,NW95}.
This representation  differs from the 
Keplerian representation of the 
Newtonian motion through the appearance of three
eccentricities instead of one, and a constant measuring the
secular advance of the periastron. 
Starting from the above representation of the orbital elements
in terms of the conserved energy and angular momentum, one
computes the time variation of the orbital elements. 
One ends up with a result, in terms of the  time variation
of the `conserved' energy and angular momentum.
By a  heuristic argument, one replaces these by the corresponding
{\em average } far-zone  fluxes which may be computed by  averaging
the far-zone fluxes over an orbital period,  using the 
quasi-Keplerian
orbital representation. 
The reduction in
the orbital period, accurately inferred from the timing data
of the binary pulsars  is in excellent agreement with the rigorous
predictions of general relativity\cite{HT75,DD81,TD83a,TD83b,Wol94},
which in turn are consistent with the results of the above heuristic
approach\cite{PM63,EH75,W75}.
Extending the above approach, Blanchet and Sch\"afer have  obtained
the 1PN and the 1.5PN corrections to $\dot P $, the
rate of decay of the orbital period $P$
\cite{BS89,BS93}.
They have shown that for PSR 1913+16, the
relative 1PN and 1.5PN corrections are numerically equal to
$ + 2.15\,\times 10^{-5}$ and $ +1.65\,\times 10^{-7}$ respectively.
These are unfortunately
far below the present accuracy in the measurements of $\dot P$
for 1913+16. 
Junker and Sch\"afer\cite{JS92}  computed the 1PN contributions to  the 
gravitational waveforms, the  
associated angular and linear 
momentum fluxes and used it to compute the 
evolution of the orbital elements in
the quasi-Keplerian representation.
In the other part of the paper, we extend the above computations
to obtain the 2PN corrections to the  evolution of orbital
elements, taking due care of  a new complication at this order
that the far-zone fluxes are computed in the harmonic
or De Donder coordinates, while the orbital representation is
available in the Arnowit, Deser and Misner (ADM)coordinates.

Briefly,  in this paper  we obtain  the 
terms O($\epsilon^2$) in the expressions below,
where  $\epsilon \sim v^2/c^2  \approx Gm / c^2\,r$;
$m, r, v $,  being the total mass, the distance between the bodies and
the relative velocity of the two bodies, respectively,
\begin{mathletters}
\begin{eqnarray}
I_{ij} &=& (I_{ij})_{\rm N} 
\left \{ 1 +  O(\epsilon)
+ O(\epsilon^2)+ ...\right \}\,,\\
h^{TT}_{km} &=& ( h^{TT}_{km})_{\rm N}
\left \{ 1 + O( \epsilon^{0.5}) + O(\epsilon)
+ O(\epsilon^{1.5}) + O(\epsilon^2)+ ...\right \}\,,\\
{d{\cal E} \over dt } &=& ({d{\cal E} \over dt })_{\rm N} \left \{ 1 
+ O(\epsilon) 
+ O(\epsilon^{1.5}) + O(\epsilon^2) +..\right \} \,,\\
{d{\bf {\cal J}} \over dt }&=& ({d{\bf {\cal J}} \over dt })_{\rm N}
\left \{ 1 + O(\epsilon)
+O(\epsilon^{1.5}) + O(\epsilon^2) +...\right \}\,,\\
< {d {\cal E}\over dt} >&=& <{d {\cal E}\over dt}>_{\rm N}
\left \{ 1 + O(\epsilon) + O(\epsilon^{1.5}) 
+ O(\epsilon^2) +..\right \} \,,\\
< {d {\cal J}\over dt} >&=& <{d {\cal J}\over dt}>_{\rm N} 
\left \{ 1 + O(\epsilon) + O(\epsilon^{1.5}) 
+ O(\epsilon^2) +..\right \} \,,\\
< {d a_r\over dt} >&=& <{d a_r\over dt}>_{\rm N}
\left \{ 1 + O(\epsilon) + O(\epsilon^{1.5})
+ O(\epsilon^2) +..\right \} \,,\\
< {d e_r\over dt} >&=& <{d e_r\over dt}>_{\rm N}
\left \{ 1 + O(\epsilon) + O(\epsilon^{1.5})
+ O(\epsilon^2) +..\right \} \,,\\
{dP \over dt} &=& ({dP \over dt})_{\rm N}\left \{ 1 + O(\epsilon)
+ O(\epsilon^{1.5}) + O(\epsilon^2) +...\right \}\,,
\end{eqnarray}
\label{aim}
\end{mathletters}
and where
$ I_{ij} $ is the mass quadrupole moment for a system of two
compact objects moving in  general orbits while 
$ h^{TT}_{km}$ is the transverse-traceless(TT) part of the 
radiation-field, representing the deviation of the metric
from the flat spacetime. In the above, d${\cal E}$/dt 
, d${\bf {\cal J}}$/dt are the 
far-zone energy and angular momentum fluxes,  
$< {d {\cal E} / dt} >$ and $ < {d {\cal J}/ dt} > $ represent the averages
of the far-zone fluxes over an orbital period, while  $<da_r/dt>$, 
 $<de_r/dt >$   
along with $dP/dt$  give 
the gravitational radiation driven
rate of decay of the orbital elements  of the
binary in the generalized quasi- Keplerian parameterization. 
Note that the suffix `${\rm N }$'  denotes Newtonian
contribution in all the above equations. For example 
$(h^{TT}_{km} )_{\rm N}$ denotes the Newtonian 
contribution to the waveform given by 
$ \left \{ { 2\,G/ (c^4\,R)}\right \}{\cal P}_{ijkm}\,I^{(2)}_{ij} $. 
See Eq. (\ref{wff}) for our notation.

                       The plan of the paper is as follows: 
In section~\ref{sec:moments}, using the BDI approach,
we compute the 2PN
accurate mass quadrupole moment for two masses moving on  general orbits.  
We also  obtain and list 
the mass and the current moments to the required accuracy, 
needed to compute the 2PN accurate waveform. 
In section III we calculate  
the 2PN contributions to the far-zone 
energy and angular momentum fluxes and discuss the limiting forms
of these expressions.
In section~\ref{sec:dpdt2}, for the quasi-elliptic case,  we average the above expressions
over an orbital period to  obtain the 2PN  
corrections to  $< {d{\cal E}/dt}>$, $<{d{\cal J}/ dt}>$ and 
the rate of decay of the orbital elements.
Section~\ref{sec:wf2} computes the 2PN contribution to
gravitational waveform for general orbits.
Section~\ref{sec:conc} contains the summary and a few 
concluding remarks.
In Appendix A we list  identities, that are used  
in the computations, especially, of the waveform.
Finally, in appendix B, we sketch the steps involved in 
verifying the equivalence of our waveform obtained using
the STF multipoles of the radiative field and the Will-Wiseman
one obtained using the Epstein Wagoner multipoles.

\section{Mass and current moments of compact binaries on general orbits
for 2PN generation} 
\label{sec:moments}
\subsection{2PN  mass quadrupole moment}
The starting point for the computation of the 2PN accurate mass moment is
the form of the moment quoted in \cite{BDI95} i.e., Eq.(2.17).
\begin{eqnarray}
 I_L (t) &=& {\rm FP}_{B=0} \int d^3{\bf x} |{\bf x}|^B \left\{ \hat x_L
 \left[\sigma -{4\over c^4}\sigma U_{ss} +{4\over c^4} U\sigma_{ss}\right]
 + {|{\bf x}|^2\hat x_L\over 2c^2(2\ell+3)} \partial^2_t \sigma
    \right.\nonumber\\
 &&- {4(2\ell+1) \hat x_{iL}\over c^2(\ell+1)(2\ell+3)} \partial_t
 \left[ \left( 1+{2U\over c^2}\right) \sigma_i -{2U_i\over c^2}\sigma
 + {1\over \pi Gc^2} \left( \partial_jU \partial_i U_j - {3\over 4}
  \partial_i U\partial_j U_j\right) \right]\nonumber \\
 &&+ {|{\bf x}|^4 \hat x_L\over 8c^4(2\ell+3)(2\ell+5)} \partial^4_t
 \sigma -{2(2\ell+1)|{\bf x}|^2\hat x_{iL}\over c^4(\ell+1)(2\ell+3)(2\ell+5)}
  \partial^3_t \sigma_i \nonumber \\
 &&+ {2(2\ell+1) \hat x_{ijL}\over c^4(\ell+1)(\ell+2)(2\ell+5)}
\partial^2_t \left[\sigma_{ij}+{1\over 4\pi G} \partial_i U\partial_j U\right]
\nonumber \\
 &&\left.+ {\hat x_L\over \pi Gc^4} \left[ 2U_i \partial_{ij} U_j
 - U_{ij} \partial_{ij} U - {1\over 2}(\partial_i U_i)^2
  +2\partial_i U_j \partial_j U_i - {1\over 2} \partial^2_t
 (U^2) + W_{ij} \partial_{ij} U \right] \right\}\ +
O(\varepsilon^5) 
\label{eq:2.17}
\end{eqnarray}
 The symbol ${\rm FP}_{B=0}$ in the above  stands for ``Finite
Part at $B=0$'' and denotes a mathematically well-defined operation of
analytic continuation. For more details see \cite{BDI95}.

As emphasized in \cite{BDI95} though the above expression is
mathematically well defined, it is a non-trivial and long calculation to
rewrite it explicitly in terms of the source variables only. 
This is achieved
by representing the stress energy tensor of the source as a sum of Dirac
$\delta$-functions. 
\begin{equation}
 T^{\mu\nu} ({\bf x},t) = \sum^N_{A=1} m_A {dy^\mu_A\over dt}
 {dy^\nu_A\over dt} {1\over \sqrt{-g}} {dt\over d\tau} \delta ({\bf x}
 - {\bf y}_A (t))\ , \label{eq:2.18}
\end{equation}
where $m_A$ denotes the (constant) Schwarzschild mass of the $A$th
compact body.
Evaluating this to 2PN accuracy we obtain for the source variables
\begin{mathletters}
\label{eq:2.19}
\begin{eqnarray}
 \sigma ({\bf x},t) &=& \sum_{A=1}^N \mu_A (t) \left( 1+ {{\bf v}^2_A
 \over c^2} \right) \delta ({\bf x} -{\bf y}_A (t))\ , \label{eq:2.19a}\\
 \sigma_i ({\bf x},t) &=& \sum_{A=1}^N \mu_A (t) v^i_A \delta
  ({\bf x} -{\bf y}_A (t))\ , \label{eq:2.19b}\\
 \sigma_{ij} ({\bf x},t) &=& \sum_{A=1}^N \mu_A (t) v^i_A v^j_A \delta
  ({\bf x} -{\bf y}_A (t))\ , \label{eq:2.19c}
\end{eqnarray}
\end{mathletters}
where $v^i_A \equiv dy^i_A /dt$ and
\begin{mathletters}
\label{eq:2.20}
\begin{eqnarray}
 \mu_A (t)&=&m_A \biggl \{1 +(d_2)_A+(d_4)_A\biggr \}\ ,\label{eq:2.20a}\\
 d_2 &\equiv& {1\over c^2} \left \{ {1 \over 2}{\bf v}^2 -V \right\}\ ,
\label{eq:2.20b}\\
 d_4 &\equiv& {1\over c^4} \left\{ {3\over 8} {\bf v}^4 + {3\over 2}
  U{\bf v}^2 - 4 U_iv_i - 2 \Phi + {3\over 2} U^2 + 4 U_{ss} \right\}\ ,
  \label{eq:2.20c}
\end{eqnarray}
\end{mathletters}
In the above  $V$ denotes the combination
\begin{equation}
 V \equiv U + {1\over 2c^2}\, \partial_t^2 X\ , \label{eq:2.21}
\end{equation}
the potential appearing naturally in the 1PN near-zone metric
in harmonic coordinates. The subscript $A$ appearing in Eq.~(\ref{eq:2.20a})
indicates that one must replace the field point ${\bf x}$ by the position
${\bf y}_A$ of the $A$th mass point, while discarding all the ill-defined
(formally infinite) terms arising in the limit ${\bf x}\to {\bf y}_A$. For
instance
\begin{mathletters}
\begin{eqnarray}
 (U)_A &=& G \sum_{B\not= A} {\mu_B(t) (1+{\bf v}^2_B /c^2)\over
    |{\bf y}_A - {\bf y}_B|} \,, \\
 (U_{ss})_A &=& G \sum_{B\not= A} {\mu_B(t) {\bf v}^2_B \over
    |{\bf y}_A - {\bf y}_B|} \,, \\
 (\Phi)_A &=& G \sum_{B\not= A} {\mu_B(t) (1+{\bf v}^2_B /c^2)(U)_B
\over |{\bf y}_A - {\bf y}_B|} \,, \\
 (X)_A &=& G \sum_{B\not= A} {\mu_B(t) (1+{\bf v}^2_B /c^2)
    |{\bf y}_A - {\bf y}_B|} \ . 
\end{eqnarray}
\end{mathletters}
[Note that the second time derivative appearing in $V$, Eq.~(\ref{eq:2.21}),
must be explicated before making the replacement ${\bf x}\to {\bf y}_A (t)$.]

The terms in Eq.(\ref{eq:2.17}) fall into 3  types: 
compact terms, $Y$~terms  and $W$ terms.
The compact terms, where the 3-dimensional integral extends only over the
compact support of the material sources; the $Y$ terms involving
three dimensional integral of the product of two Newtonian like
potentials; and the W term involving three dimensional integrals of terms
trilinear in source variables. The evaluation of these different terms
proceeds exactly as in the circular case. In fact if the time derivatives
are not explicitly implemented the expression in the general case and the
circular case would be identical. The difference obtains when the time
derivatives are implemented using the equation of motion. In this section
we need to use the  general form of the Damour-Deruelle 
equations of motion
rather than the restricted form of the circular orbit equations of motion
relevant in \cite{BDI95}.

We take up the compact terms first. They are given by
\begin{eqnarray}
 I^{(C)}_L &=& \sum^N_{A=1} \left\{ \tilde\mu_A \left[ 1 -{4\over c^4}
   U^A_{ss} + {4\over c^4} U^A ({\bf v}_A)^2\right] \hat y^L_A
   \right. \nonumber\\
 && + {1\over 2(2\ell+3)c^2} {d^2\over dt^2} (\tilde\mu_A {\bf y}_A^2
  \hat y_A^L) + {1\over 8(2\ell+3)(2\ell+5)c^4} {d^4\over dt^4}
 (\tilde\mu_A ({\bf y}^2_A)^2 \hat y^L_A) \nonumber \\
 && -{4(2\ell+1)\over (\ell+1)(2\ell+3)c^2} {d\over dt} \left(
  \left[ \mu_A \left( 1+ {2U^A\over c^2} \right) v^i_A - {2U^A_i\over c^2}
  \tilde\mu_A \right] \hat y^{iL}_A\right)\nonumber \\
 && -{2(2\ell+1)\over (\ell+1)(2\ell+3)(2\ell+5)c^4} {d^3\over dt^3}
   (\mu_A v^i_A {\bf y}^2_A \hat y^{iL}_A ) \nonumber \\
 &&\left. +{2(2\ell+1)\over (\ell+1)(\ell+2)(2\ell+5)c^4} {d^2\over dt^2}
   (\mu_A v^i_A v^j_A \hat y^{ijL}_A ) \right\}\ , \label{eq:3.5}
\end{eqnarray}
in which we have introduced for convenience $\tilde\mu_A \equiv \mu_A
(1+{\bf v}^2_A/c^2).$  
In the above form the moment depends not only on the position and velocity
of the bodies but on higher time derivatives. It is in the reduction of
these derivatives that we need the 2PN accurate equation of motion for
general orbits. We use a harmonic coordinate system in which 
the 2PN center of
mass is at rest at the origin. Using the  2PN accurate center of mass theorem,
in the center of mass frame,  we can express the individual positions 
of the two bodies moving in general orbits 
in terms of their relative position ${\bf x}={\bf y{_1}}-{\bf y_{2}}$
and velocity  ${\bf v}={\bf v_{1}}-{\bf v_{2}}$
\begin{mathletters}
\begin{eqnarray}
{\bf y_{1}}&=&\left\{X_2+\frac{\eta\delta m}{2mc^2}\left[v^2-
\frac{Gm}{r}\right]
+\frac{\chi_1}{c^4}\right\}{\bf x}+
\frac{\chi_2}{c^4}{\bf v}\,, \\
{\bf y_{2}}&=&\left \{ -X_1+\frac{\eta\delta m}{2mc^2}\left[v^2
-\frac{Gm}{r}\right]
+\frac{\chi_1}{c^4}\right \}{\bf x}+\frac{\chi_2}{c^4}{\bf v}\,,
\end{eqnarray}
\label{ddcom}
\end{mathletters}

where $r=|{\bf y_{1}}-{\bf y_{2}}|$
is the harmonic separation between the 2 bodies. 
The explicit values of
$\chi_1$ and $\chi_2$ are not needed in our calculations and hence not
given above. 
The above equations are obtained by setting equal to zero the conserved
mass dipole {\bf G} for general orbits. Here we denote
\begin{eqnarray}
 m \equiv m_1 + m_2\ ,\quad \delta m \equiv m_1-m_2\ , \quad 
\nonumber \\
 X_1 \equiv \frac{m_1}{m}\ , \quad 
X_2\equiv \frac{m_2}{m}\ =1-X_1 , \nonumber \\
\eta \equiv X_1X_2\ = \frac{m_1m_2}{m^2} \equiv  \frac{\mu}{m}\,.
\end{eqnarray} 

     The 2PN accurate equations of motion  is
written down next for completeness, where finite-size effects, 
such as spin-orbit,
spin-spin, or tidal interactions are ignored. 
\cite{dd81,grishchuk,LW90}. For the relative motion we have
\begin{equation}
{\bf a} = {\bf a}_N + {\bf a}_{\rm PN}^{(1)} +
{\bf a}_{\rm 2PN}^{(2)}
+ O(\epsilon^5) \,,
\label{aPNgeneral}
\end{equation}
where the subscripts denote the nature of the term,
Newtonian~(N), post-Newtonian
(PN), post-post-Newtonian (2PN),
and the superscripts denote the order in $\epsilon$.
The explicit expressions for various terms mentioned above
are given by 
\begin{mathletters}
\label{aPN}
\begin{eqnarray}
{\bf a}_N = && - {Gm \over r^2} {\bf n} \,, \label{aN}
\\
{\bf a}_{\rm PN}^{(1)} = && - {G m \over c^2\,r^2}
 \biggl\{   \left[
-2(2+\eta)
{Gm \over r} + (1+3\eta)v^2 - {3 \over 2} \eta \dot r^2 \right] {\bf n}
  -2(2-\eta) \dot r {\bf v} \biggr\} \,, \label{a1PN} \\
{\bf a}_{\rm 2PN}^{(2)} = && - {G\,m \over c^4\, r^2} 
\biggl\{  \biggl[ {3
\over 4}
(12+29\eta) {G^2\,m^2 \over r^2} + \eta(3-4\eta)v^4 + {15 \over 8}
\eta(1-3\eta)
\dot r^4 \nonumber \\
&& - {3 \over 2} \eta(3-4\eta)v^2 \dot r^2
- {1 \over 2} \eta(13-4\eta) {G\,m \over r} v^2 - (2+25\eta+2\eta^2)
{G\,m \over r} \dot r^2 \biggr]{\bf n} \nonumber \\
 && - {1 \over 2}  \left[ \eta(15+4\eta)v^2 -
(4+41\eta+8\eta^2)
{G\,m \over r} -3\eta(3+2\eta) \dot r^2 \right]\dot r {\bf v} \biggr\} \,,
\label{a2PN}
\end{eqnarray}
\end{mathletters}
where $ {\bf n}= {\bf x}/r$ and $ \dot r = dr /dt $.

        We have on hand all the ingredients to
 compute $I_c$. Though long and
tedious the computation is straightforward and yields for the 2PN mass
quadrupole: 
\begin{eqnarray}
\label{m2C}
I_{ij}^{ [C]}&=&\eta\,m\,{\rm STF}_{ij}\left \{ x^{ij}+\right.\nonumber\\
&&+\frac{1}{42c^2}\left\{x^{ij}\left[29(1-3\eta)v^2-6(5-8\eta)\frac{Gm}{r}
\right]\right.\nonumber\\
&&\left.-24(1-3\eta)r\dot{r}x^iv^j+22(1-3\eta)r^2v^{ij}\right\}
+\frac{1}{1512c^4}\left[v^4(759-5505\eta+10635\eta^2)\right.\nonumber\\
&&+\frac{G^2m^2}{r^2}(1758-6468\eta+1878\eta^2)\nonumber\\
&&+v^2\frac{Gm}{r}(5818-16742\eta-12166\eta^2)\nonumber\\
&&\left.-\dot{r}^2\frac{Gm}{r}(2038-6662\eta+146\eta^2)\right]x^{ij}\nonumber\\
&& + \frac{1}{378c^4}\left[v^2(123-1011\eta+2199\eta^2)\right.\nonumber\\
&&+\frac{Gm}{r}(68+434\eta-2090\eta^2)\nonumber\\
&&\left.+30\dot{r}^2(1-5\eta+5\eta^2)\right]r^2v^{ij}\nonumber\\
&&-\frac{1}{378c^4}\left[\frac{Gm}{r}(101+287\eta-1655\eta^2)\right.\nonumber\\
&&\left.\left.+v^2(156-1212\eta+2508\eta^2)
\right]r\dot{r}x^iv^j \right\}
\end{eqnarray}

The $Y$ terms on the other hand are given by
\begin{eqnarray}
I_{ij}^{[Y]}&=&-\frac{2Gm_1m_2}{c^4}\left\{2Y^{ij}_{v_1v_2}-Y^{ij}_{v_1v_1}
\right.\nonumber\\
&&-\frac{1}{2}{}_{v_1}Y^{ij}_{v_2}+2{}_{v_2}Y^{ij}_{v_1}-\frac{1}{2}
\partial_t^2(Y^{ij})\nonumber\\
&&-\frac{20}{21}\partial_t\left[{}_{v_2}Y^{aij}_{a}-
\frac{3}{4}{}_{v_1}Y^{aij}_{a}\right]\nonumber\\
&&\left.+\frac{5}{216}\partial_t^2\left[{}_{a}Y^{abij}_{b}\right]\right\}
+(1\leftrightarrow 2)
\end{eqnarray}
where following \cite{BDI95}
\begin{mathletters}
\begin{eqnarray}
{}_{v_1}Y_{v_{2}}^{L}&=& v_{1}^{a}\,v_{2}^{b}\, {}_a Y_b^{L}\,,\\
{}_a Y_b^{L} &=& \partial_{ y_1^a}\, \partial_{y_2^b} Y^{L}\,,\\
Y^L ( {\bf y_{1}},{\bf y_{2}}) &=& {|{\bf y_{1}}-{\bf y_{2}}|\over l+1}
\sum_{p=0}^{l} y_1^{<l-p} y_2^{p>} \,,
\end{eqnarray}
\end{mathletters}

\begin{mathletters}
so that
\begin{eqnarray}
Y^{ij}_{v_1v_2}&\equiv&2v_1^sv_2^kY^{ij}_{sk}\\
Y^{ij}_{sk}&=&\frac{1}{3}\frac{\partial}{\partial y_2^s}
\frac{\partial}{\partial y_2^k}r_{12}\left(y_1^{ij}+y_1^{(i}y_2^{j)}+
y_2^{ij}\right)\\
r_{12}&=&|{\bf y_{1}}-{\bf y_{2}}|
\end{eqnarray}
\end{mathletters}

The explication of all the above terms finally leads us to
\begin{eqnarray}
\label{m2Y}
I_{ij}^{[Y]}&=&-\frac{2\,m\,\eta}{63\,c^4}\,\frac{Gm}{r}\,
{\rm STF}_{ij}\left\{ x^{ij}
\left[(v^2-\dot{r}^2)(37-101\eta-50\eta^2)\right.\right.\nonumber\\
&&\left.+\frac{Gm}{r}(18-54\eta-3\eta^2)\right]\nonumber\\
&&-r^2v^{ij}(118-92\eta+10\eta^2)\nonumber\\
&&\left.+r\dot{r}x^iv^j(82-362\eta+16\eta^2)\right\}
\end{eqnarray}

The evaluation of the $I^{[W]}$ term, the new feature at 2PN level, was
discussed in detail in \cite{BDI95}. The $W$ term has been evaluated there
for general orbits and we need to use the same result here. We have
\begin{eqnarray}
\label{m2W}
I^{[W]}_{ij}&=&-\frac{\eta\, m}{c^4}\,\frac{G^2\,m^2}{r^2}
{\rm STF}_{ij}\left \{[ 2+5\eta]x^{ij} \right \}
\end{eqnarray}
Adding up the compact  {\em i.e.,} $C$, $Y$ and $W$ contributions 
given by Eqs.(\ref{m2C}), (\ref{m2Y}) and (\ref{m2W}), we
finally obtain the expression for the 2PN accurate mass quadrupole for a
system of two bodies moving in general orbits. The final result is
written below as a combination of the three possible combinations
$x^{ij},x^iv^j,v^{ij}$ with coefficients which include corrections
beyond the Newtonian order at 1PN and 2PN orders:
\begin{eqnarray}
\label{m22}
I_{ij} &=& \mu\, {\rm STF}_{ij}\biggl \{ x^{ij}
        \biggl [1 +\nonumber \\
       &+& \frac{1}{42\,c^2}\;\left( (29-87\eta)v^2 -
           (30-48\eta)\frac{Gm}{r}\right )\nonumber \\
       &+& \frac{1}{c^4}\left (\frac{1}{504} (253-1835\eta +
           3545\eta^2) v^4\right.\nonumber \\
       &+& \left.\frac{1}{756} (2021-5947\eta -4883\eta^2)
           \frac{Gm}{r}\,v^2\right.\nonumber \\
       &-& \left.\frac{1}{756} (131-907\eta +1273\eta^2)
           \frac{Gm}{r}\,\dot{r}^{2}\right.\nonumber \\
       &-& \left.\frac{1}{252} (355+1906\eta-337\eta^2)
           \frac{G^2m^2}{r^2}\right )\biggr ]\nonumber \\
       &-& x^{i}v^{j}\left[\frac{r\dot{r}}{42\,c^2}
         (24-72\eta)\right.\nonumber\\
       &+& \left.\frac{r\dot{r}}{c^4} \left( \frac{1}{63}
           (26-202\eta +418\eta^2) v^2\right.\right.\nonumber \\
       &+& \left.\left.\frac{1}{378} (1085-4057\eta -1463\eta^2)
           \frac{Gm}{r}\right )\right] \nonumber \\
       &+& v^{ij}\left[ \frac{r^2}{21\,c^2} 
       (11-33\eta)\right.\nonumber\\
       &+& \left.\frac{r^2}{c^4} \left (\frac{1}{126} (41-337\eta
           +733\eta^2)v^2\right.\right.\nonumber \\
       &+&\left.\left.\frac{5}{63} (1-5\eta
           +5\eta^2)\dot{r}^2\right.\right.\nonumber \\
       &+& \left.\left. \frac{1}{189}\, (742-335\eta -985\eta^2)
           \frac{Gm}{r}\right )\right] \biggr \}
\end{eqnarray}
The above expression is identical to the one obtained by Will
and Wiseman in the appendix E of \cite{WW96} using the new 
improved version of the Epstein-Wagoner formalism. In their
treatment  the Epstein-Wagoner multipoles  appear more naturally,
using which they compute the STF mass quadrupole moment. Since the
approach employed here and in \cite{WW96} follow algebraically
different routes, the above match provides a valuable check
on the long and complicated algebra involved in the determination
of the crucial mass quadrupole moment for 2PN generation.

\subsection{The other relevant mass and current moments} 
     In this section we list the higher order mass and current 
multipole moments, required to compute the 
2PN contributions 
to the gravitational waveform and the associated far-zone 
energy and angular momentum fluxes.
They are straightforwardly obtained by explicating the point
particle limits of the more general expressions in the earlier
BDI papers\cite{bdigen}
\begin{eqnarray}
I_{ijk} &=& -(\mu\, \frac{\delta m}{m} )
          {\rm  STF}_{ijk}\left\{ \right.\nonumber \\
       && x^{ijk}\,\left.\left[1 + \frac{1}{6\,c^2} 
           \left ((5-19\eta )v^2 \right.\right.\right.\nonumber \\
       &&  \left.\left.\left.
           -(5-13\eta ) \frac{Gm}{r}\right )\right] \right.\nonumber \\
       &-& \left.x^{ij}v^k\left [\frac{r\dot{r}}{c^2}
          (1-2\eta )\right ]
       \right.\nonumber \\
       &+& \left.x^iv^{jk}\left [\frac{r^2}{c^2} 
       (1-2\eta )\right ]\right \}
\end{eqnarray}
\begin{eqnarray}
I_{ijkl} &=& \mu \,{\rm STF}_{ijkl}
             \left\{\right.\nonumber \\
        && x^{ijkl}\,\left.\left[(1-3\eta ) 
            \right. \right. \nonumber \\
        &&  \left.\left. 
            + \frac{1}{110\,c^2} 
	     \left( (103-735\eta
            +1395\eta^2) v^2\right.\right.\right.\nonumber \\
        &-& \left.\left.\left.(100-610\eta +1050\eta^2)
            \frac{Gm}{r}\right )\right]\right.\nonumber\\
        &-& \left.v^ix^{jkl}\left \{\frac{72\,r\,{\dot r}}{55\,c^2} 
	     (1-5\eta +5\eta^2) \right \}
            \right.\nonumber \\
        &+& \left.v^{ij}x^{kl}\left \{\frac{78\,r^2}{55\,c^2}
	(1-5\eta +5\eta^2) \right \}\right\}
\end{eqnarray}
\begin{eqnarray}
I_{ijklm} &=& -(\mu \,\frac{\delta m}{m})\,( 1-2\eta)
             {\rm STF}_{ijklm}\left \{ x^{ijklm}\right \}  \\
I_{ijklmn} &=& \mu (1-5\eta +5\eta^2 )
            {\rm STF}_{ijklmn}\left \{  x^{ijklmn}\right \} \\
J_{ij} &=& -(\mu\,\frac{\delta m}{m})
             {\rm  STF}_{ij} \epsilon_{jab} \left\{ 
                 \right.\nonumber \\
       && x^{ia}v^b\,\left.\left[1 + \frac{1}{28\,c^2}
           \left ((13-68\eta )v^2 
        \right. \right.\right. \nonumber \\
       && \left.\left.\left. 
        + (54+60\eta )\frac{Gm}{r}\right )\right] \right.\nonumber \\
       &+& \left.v^{ib}x^a \left [\frac{r\,{\dot r}}{28\,c^2} (5-10\eta )
            \right ] \right\}
\end{eqnarray}
\begin{eqnarray}
J_{ijk} &=& \mu \,{\rm STF}_{ijk}
            \epsilon_{kab}\left\{ x^{aij}v^b
	    \biggl [ (1-3\eta )\right.\nonumber \\
        &+& \left.\frac{1}{90\,c^2}\left ((41-385\eta
            +925\eta^2)v^2\right.\right.\nonumber \\
        &+& \left.\left.(140-160\eta
            -860\eta^2)\frac{Gm}{r}\right )\biggr ] \right.\nonumber \\
        &+& \left.\frac{7\,r^2}{45\,c^2}
	(1-5\eta +5\eta^2)\, x^av^{ijb}\right.\nonumber \\
        &+& \left.\frac{10\,r\,{\dot r}}{45\,c^2} (1-5\eta +5\eta^2) 
            x^{ai}v^{bj}\right \}
\end{eqnarray}
\begin{eqnarray}
J_{ijkl} &=& -\left(\mu\,\frac{\delta m}{m} (1-2\eta )\right )
            {\rm STF}_{ijkl}\left \{
            \epsilon_{lab}\, x^{aijk}v^b \right \} \\
J_{ijklm} &=& \left ( \mu\, (1-5\eta +5\eta^2)\right ) \,
             {\rm STF}_{ijklm}\left \{
            \epsilon_{mab}\, x^{aijkl}v^b\right \}
\end{eqnarray}
The mass and the current moments listed above, agree with Eqs. (E3) of
\cite{WW96}.
For the case of circular orbits, the above mass and current moments 
 reduce to  Eqs. (4.4) of \cite{BDI95}.  
\section{The far-zone  fluxes}
\subsection{ The Energy flux} \label{sec:ef2}
As discussed in \cite{BDI95} the end result of the 2PN accurate 
generation formalism is an expression relating the radiative 
mass and current multipole moments $U_L$ and $V_L$ respectively to the
source mass and current multipole moments $I_L$ and $J_L$ respectively,
obtained in the previous section. In particular,
\begin{mathletters}
\label{eq:2.6}
\begin{eqnarray}
 U_{ij} (T_R) &=& I^{(2)}_{ij} (T_R) + {2Gm\over c^3} \int^{+\infty}_0 d\tau
  \left[ \ln \left({\tau\over 2b}\right) + {11\over 12}\right] I^{(4)}_{ij}
   (T_R-\tau) + O(\varepsilon^5)\ , \label{eq:2.6a} \\
 U_{ijk} (T_R) &=& I^{(3)}_{ijk} (T_R) + {2Gm\over c^3} \int^{+\infty}_0 d\tau
  \left[ \ln \left({\tau\over 2b}\right) + {97\over 60}\right] I^{(5)}_{ijk}
   (T_R-\tau) + O(\varepsilon^5)\ , \label{eq:2.6b} \\
 V_{ij} (T_R) &=& J^{(2)}_{ij} (T_R) + {2Gm\over c^3} \int^{+\infty}_0 d\tau
  \left[ \ln \left({\tau\over 2b}\right) + {7\over 6}\right] J^{(4)}_{ij}
   (T_R-\tau) + O(\varepsilon^4)\ , \label{eq:2.6c}
\end{eqnarray}
\end{mathletters}
for the moments that need to be known beyond the 1PN accuracy, and
\begin{mathletters}
\label{eq:2.7}
\begin{eqnarray}
 U_L (T_R) &=& I^{(\ell)}_L (T_R) + O(\varepsilon^3)\ , \label{eq:2.7a}\\
 V_L (T_R) &=& J^{(\ell)}_L (T_R) + O(\varepsilon^3)\ , \label{eq:2.7b}
\end{eqnarray}
\end{mathletters}
for the other ones. The integrals in the above expressions,  
associated with the gravitational wave tails, contain in addition
to the total mass-energy of the source $m$, a quantity $b$ which is
an arbitrary constant with dimensions of time parametrizing a freedom
associated in the construction of the far-zone radiative coordinate system.

In terms of the STF radiative moments of the gravitational field
the far-zone energy flux to 2PN accuracy is given by\cite{LB96} 
(with $U^{(n)}\equiv d^nU/dT_R^n)$:
\begin{eqnarray}
 \left({d{\cal E}\over dt}\right)_{\rm far-zone}={G\over c^5}\biggl\{ 
{1\over 5} U^{(1)}_{ij} U^{(1)}_{ij}
 &+&{1\over c^2} \left[ {1\over 189} U^{(1)}_{ijk} U^{(1)}_{ijk}
 + {16\over 45} V^{(1)}_{ij} V^{(1)}_{ij}\right] \nonumber \\
 &+&{1\over c^4} \left[ {1\over 9072} U^{(1)}_{ijkm} U^{(1)}_{ijkm}
 + {1\over 84} V^{(1)}_{ijk} V^{(1)}_{ijk}\right]+O(\varepsilon^6)\biggr\}\ .
 \label{eq:2.5}
\end{eqnarray}

The 2PN-accurate energy loss given by Eq.~(\ref{eq:2.5})  splits naturally
 into an ``instantaneous" contribution and a
``tail'' one. In this paper, we  deal only with 
the instantaneous contribution,
which is given by
\cite{KT80,BDI95}
\begin{eqnarray}
\label{2pnf}
\left( {d{\cal E}\over dt}\right)_{\rm far-zone}^{\rm inst}
=  {G\over c^5} \biggl\{
  {1\over 5} I^{(3)}_{ij} I^{(3)}_{ij}
 &+&{1\over c^2} \left[ {1\over 189} I^{(4)}_{ijk} I^{(4)}_{ijk}
 + {16\over 45} J^{(3)}_{ij} J^{(3)}_{ij}\right] \nonumber \\
 &+&{1\over c^4} \left[ {1\over 9072} I^{(5)}_{ijkm} I^{(5)}_{ijkm}
 + {1\over 84} J^{(4)}_{ijk} J^{(4)}_{ijk}\right] \biggr\}\ .
 \label{oo2pnf}
\end{eqnarray}
Here \(I_{L}^{(n)}\) denotes the $n^{th}$ time 
derivative of STF multipole moment
of rank $L$. All the computations from here onwards are 
performed, using MAPLE \cite{Maple}. 
 Evaluating the   relevant time derivatives of the
multipole moments in Eq. (\ref{2pnf}),  using the post-Newtonian 
equations of motion to the appropriate order we  obtain
\begin{mathletters}
\label{gief}
\begin{eqnarray}
\biggl(\frac{d{\cal E}}{dt}\biggr )_{\rm far-zone}^{\rm inst} &=& \dot{{\cal E}}_N + 
\dot{{\cal E}}_{1PN} + \dot{{\cal E}}_{2PN}\,,\\
\dot{{\cal E}}_N &=& {8 \over 15}\frac{G^3\,m^2\,\mu^2}{c^5\,r^4}
           \left\{ 12v^2 - 11\dot{r}^2\right\} \,,\\
\dot{{\cal E}}_{1PN} &=&{8 \over 15}\frac{G^3\,m^2\,\mu^2}{c^7\,r^4}
               \left\{\frac{1}{28}
               \left[(785 - 852\eta)v^4\right.\right.\nonumber \\
              &-& \left.\left.2(1487 - 1392\eta)
	      v^2\dot{r}^2\right.\right.\nonumber \\
              &-& \left.\left.160(17
-\eta) \frac{Gm}{r}\,v^2\right.\right.\nonumber \\
              &+& \left.\left.3(687 - 620\eta)\dot{r}^4
+ 8 (367 - 15\eta)\frac{Gm}{r}\,\dot{r}^2\right.\right.\nonumber \\
              &+& \left.\left.16(1 - 4\eta)\,\frac{G^2m^2}{r^2}
\right]\right\}\,,\\
\dot{{\cal E}}_{2PN} &=&{8 \over 15}\frac{G^3\,m^2\,\mu^2}{c^9\,r^4}
               \left\{
                  \frac{1}{42} (1692 - 5497\eta
                 + 4430\eta^2)v^6\right.\nonumber \\
              &-& \left.\frac{1}{14} (1719 - 10278\eta
+ 6292\eta^2) v^4\dot{r}^2\right.\nonumber \\
              &-& \left.\frac{1}{21} (4446 - 5237\eta
+ 1393\eta^2)\frac{Gm}{r}\,v^4\right.\nonumber \\
              &+& \left.\frac{1}{14} (2018 - 15207\eta
+ 7572\eta^2)v^2\dot{r}^4\right.\nonumber \\
              &+& \left.\frac{1}{7} (4987 - 8513\eta
+ 2165\eta^2)\frac{Gm}{r}\,v^2\dot{r}^2\right.\nonumber \\
              &+& \left.\frac{1}{756} (281473 + 81828\eta
+ 4368\eta^2)\frac{G^2m^2}{r^2}\,v^2\right.\nonumber \\
              &-& \left.\frac{1}{42} (2501 - 20234\eta
+ 8404\eta^2)\dot{r}^6 \right.\nonumber \\
              &-& \left.\frac{1}{63} (33510 - 60971\eta
+ 14290\eta^2)\frac{Gm}{r}\,\dot{r}^4\right.\nonumber \\
              &-& \left.\frac{1}{252} (106319 + 9798\eta
+ 5376\eta^2) \frac{G^2m^2}{r^2}\,\dot{r}^2\right.\nonumber \\
              &+& \left.\frac{2}{63} (-253 + 1026\eta
- 56\eta^2)\frac{G^3m^3}{r^3}\right\}
\end{eqnarray}
\end{mathletters}
 Eqs. (\ref{gief})  are in exact agreement with the results of
 Will and Wiseman using the new improved Epstein-Wagoner approach\cite{WW96}.
Circular and radial infall limits 
of Eqs. (\ref{gief}) are in agreement with 
earlier results \cite {BDIWW95,BDI95,SPW96,WW96} and discussed further 
in section~\ref{sec:limits}.

The tail contribution, on the other hand is given by 
\begin{eqnarray}
 \left({d{\cal E}\over dt}\right)_{\rm far-zone}^{\rm tail}&=&{2G\over 5c^5}
 {2Gm\over c^3} I^{(3)}_{ij} (T_R) \int^{+\infty}_0 d\tau\, \ln
 \left( {\tau\over 2b_1}\right) I^{(5)}_{ij} (T_R-\tau)\,, \label{eq:4.13}
\end{eqnarray}
where $b_1 \equiv b\, e^{-11/12}$.
A detailed discussion of the tail terms and its implications
has been given by Blanchet and Sch\"afer\cite{BS93}, and we do not discuss
it any further in this paper. 

\subsection{The  angular momentum flux}
\label{sec:amf2}
In terms of the STF radiative multipole moments the far-zone angular
momentum flux to 2PN accuracy is given by:
\begin{eqnarray}
\left( {d{{\cal J}}_i\over dt }\right)_{\rm far-zone} =
{G\over c^5}\epsilon_{ipq} \biggl\{
  {2\over 5} U_{pj} U^{(1)}_{qj}
 &+&{1\over c^2} \left[ {1\over 63} U_{pjk} U^{(1)}_{qjk}
 + {32\over 45} V_{pj} V^{(1)}_{qj}\right] \nonumber \\
 &+&{1\over c^4} \left[ {1\over 2268} U_{pjkl} U^{(1)}_{qjkl}
 + {1\over 28} V_{pjk} V^{(1)}_{qjk}\right] \biggr\}\ .
\label{amf_pri}
\end{eqnarray}
As before, rewriting  the radiative moments in terms of 
the source moments, allows us to  
separate the instantaneous and tail contributions
and  we discuss them independently. We have 
\cite{KT80}
\begin{eqnarray}
\left( {d{{\cal J}}_i\over dt }\right)_{\rm far-zone}^{\rm inst} =
{G\over c^5}\epsilon_{ipq} \biggl\{
  {2\over 5} I^{(2)}_{pj} I^{(3)}_{qj}
 &+&{1\over c^2} \left[ {1\over 63} I^{(3)}_{pjk} I^{(4)}_{qjk}
 + {32\over 45} J^{(2)}_{pj} J^{(3)}_{qj}\right] \nonumber \\
 &+&{1\over c^4} \left[ {1\over 2268} I^{(4)}_{pjkl} I^{(5)}_{qjkl}
 + {1\over 28} J^{(3)}_{pjk} J^{(4)}_{qjk}\right] \biggr\}\ .
\end{eqnarray}
Computing the required time derivatives of the STF moments, using the
post-Newtonian equations of motion to the appropriate order,  we obtain  
\begin{mathletters}
\label{giamf}
\begin{eqnarray}
\biggl (\frac{d{\bf {\cal J} }}{dt}\biggr )_{\rm far-zone}^{\rm inst}
 &=& \dot {\bf {\cal J}}_N 
+ \dot {\bf {\cal J} }_{1PN} 
+ \dot {{\bf {\cal J}}}_{2PN}\,,\\
\dot{\bf {\cal J}}_N &=& 
{8 \over 5} {G^2\,m\,\mu^2 \over c^5\,r^3}{\bf {\tilde L_{N}}}
                  \left\{ 2v^2 - 3\dot{r}^2
                   + 2~\frac{Gm}{r}\right \}\,,\\
\dot{\bf {\cal J}}_{1PN} &=& {8 \over 5} 
{G^2\,m\,\mu^2 \over c^7\,r^3} {\bf {\tilde  L_N}}
                    \left\{
\frac{1}{84}\left[ (307 - 548\eta)v^4\right.\right.\nonumber \\
              &-& \left.\left.6(74 - 277\eta)v^2\dot{r}^2
- 4(58 + 95\eta)\frac{Gm}{r}\,v^2\right.\right.\nonumber \\
              &+& \left.\left.3(95 - 360\eta)\dot{r}^4
+ 2(372 + 197\eta)\frac{Gm}{r}\,\dot{r}^2\right.\right.\nonumber \\
              &-& \left.\left.2(745
- 2\eta)\frac{G^2m^2}{r^2}\right ]\right\}\,,\\
\dot{\bf {\cal J}}_{2PN} &=&{8 \over 5} {G^2\,m\,\mu^2 \over c^9\,r^3}
{\bf {\tilde L_{N}}} 
                    \left\{\frac{1}{504}
\left[(2665 - 12355\eta + 12894\eta^2)v^6\right.\right.\nonumber \\
              &-& \left.\left.3(2246 - 12653\eta
+ 15637\eta^2)v^4\dot{r}^2\right.\right.\nonumber \\
              &+& \left.\left.(165 - 491\eta
+ 4022\eta^2)\frac{Gm}{r}\,v^4\right.\right.\nonumber \\
              &+& \left.\left.3(3575 - 16805\eta
+ 15680\eta^2)v^2\dot{r}^4\right.\right.\nonumber \\
              &+& \left.\left.(21853 - 21603\eta
+ 2551\eta^2)\frac{Gm}{r}\,v^2\dot{r}^2\right.\right.\nonumber \\
              &-& \left.\left.2(10651 - 10179\eta
+ 3428\eta^2)\frac{G^2m^2}{r^2}\,v^2\right.\right.\nonumber \\
              &-& \left.\left.28(195 - 815\eta
+ 485\eta^2)\dot{r}^6\right.\right.\nonumber \\
             &-& \left.\left. ( 22312 -41\,398\,\eta + 
96\, 95\,\eta^2) \dot r^4 \,{G\,m \over r} 
              \right.\right.\nonumber \\
              &+& \left.\left.2(8436 - 25102\eta
+ 4587\eta^2)\frac{G^2m^2}{r^2}\,\dot{r}^2\right]\right.\nonumber \\
              &+& \left.\frac{1}{2268} (170362 + 70461\eta
+ 1386\eta^2)\frac{G^3m^3}{r^3}\right\}\,,
\end{eqnarray}
\end{mathletters}
where $ {\bf {\tilde L_N}} = {\bf r}\times {\bf v} $.
The 1PN contribution is in agreement with the earlier results 
of  Junker and Sch\"afer\cite{JS92}.
The 2PN contribution is new and together with the energy flux
obtained in the earlier section forms the starting point for
the computation the 2PN radiation reaction for compact binary
systems --4.5PN terms in the equation of 
motion -- \cite{GII97}, using the refined balance method 
proposed by Iyer and Will
\cite{IW93,IW95}.
The tail terms, in the angular momentum flux are given by
\begin{eqnarray}
 \left({d{\cal J}_i\over dt}\right)_{\rm far-zone}^{\rm tail}&=&{2G\over 5c^5}
 {2Gm\over c^3} \,\epsilon_{ijk}\,I^{(3)}_{kl} (T_R)
 \int^{+\infty}_0 d\tau\, \ln
 \left( {\tau\over 2b_1}\right) I^{(4)}_{jl} (T_R-\tau)\ .
\end{eqnarray}
We will not be discussing the tails terms here, as 
they are  extensively 
studied by Rieth and
Sch\"afer\cite{RS97}.
\subsection{ Limits}
\label{sec:limits}
All the complicated formulae, discussed in the earlier sections take
more simpler forms for quasi-circular orbits.
For compact binaries like PSR 1913+16, quasi-circular
orbits should provide a  good description
close to the inspiral phase, since   gravitational radiation reaction
would have reduced the present eccentricity, 
to vanishingly small values.
In this  context `quasi' implies the slow inspiral caused by the
radiation
reaction. The quasicircular orbit is characterised by
${\ddot r }= {\dot r } =O(\epsilon^{2.5}) $. The 2PN equations
of motion, become
\begin{mathletters}
\label{ciacc}
\begin{eqnarray}
{\bf a} &\equiv & {d{\bf v}\over dt}\equiv {d^2{\bf x}\over dt^2}
= -\omega^2_{\rm 2PN}\, {\bf x} + O(\epsilon^{2.5})\ .
\end{eqnarray}
\end{mathletters}
with  $ \omega_{\rm 2PN} $, the 2PN accurate
orbital frequency, is  given by
\begin{equation}
\omega^2_{\rm 2PN} \equiv {Gm\over r^3} \left\{ 1 -
(3-\eta){ \gamma }+
\left( 6 + {41\over 4}\eta +
\eta^2\right) {\gamma^2 }\right \}\ ,
\label{2pnomg}
\end{equation}
where $\gamma = {G\,m / c^2\,r}$.
Note that Eqs. (\ref{ciacc}) imply as usual, that
$ v \equiv |{\bf v}| = \omega_{\rm 2PN}\,r + O(\epsilon^{2.5})$, so
that from Eq. (\ref{2pnomg}) we get
\begin{equation}
{v^2} = {Gm \over r}\left\{ 1 - (3-\eta){ \gamma }+ \left( 6
+ {41\over 4}\eta + \eta^2\right) {\gamma^2} \right \}\ .
\label{eq:3.13}
\end{equation}
Substituting $\dot{r}=0$  in Eqs. (\ref{gief}),  and using Eq.
(\ref{eq:3.13})
we obtain the 2PN corrections to the far-zone energy flux
for compact binaries of arbitrary mass ratio, moving
in a quasicircular orbit
\begin{equation}
\left({d{\cal E}\over dt}\right)_{\rm far-zone}^{\rm inst}
=\frac{32}{5}\frac{c^5}{G}
\eta^2\gamma^5
\left\{1\;-\;{\gamma }\left( {2927\over 336} + {5\over
4}\eta \right) +
{\gamma^2} \left( {293383\over 9072} + {380\over 9}\eta
\right)
\right\}\,.
\label{bdief}
\end{equation}
Eq. (\ref{bdief}) is consistent with results of
\cite{BDIWW95,BDI95,WW96}.

            The energy and angular momentum 
fluxes are not independent
but related in the case of circular orbits. The precise
relation may be written  following \cite{nishi} as:
\begin{mathletters}
\label{hcirl}
\begin{eqnarray}
\biggl ({{d{\cal E}} \over {dt}} \biggr
)_{\rm far-zone}&=& v^2~\dot{  {\cal J}},\\
\mbox{ where }  \biggl ({{d {\bf {\cal J}}}
\over {dt}} \biggr
)_{\rm far-zone} &\equiv& {\bf {\tilde L_N}} \dot { {\cal J}}\,,
\end{eqnarray}
\end{mathletters}
where $v^2$ defined in terms of ${G\,m /r}$, is
given by Eq. ({\ref{eq:3.13}})\cite{BDI95}.

            The other limiting case we compare to corresponds to
the case of radial infall of
two compact objects of comparable  masses. 
Equations representing the head-on infall can be obtained
from the expressions for the  general
orbit by imposing the restrictions,
$ {\bf x}= z\hat{\bf n}$, ${\bf v}=\dot{z}\hat{\bf n}$,
$r=z$ and $v=\dot{r} = \dot{z} $.
We  consider two different cases, following
Simone, Poisson, and Will \cite{SPW96}.
In case (A),
the radial infall proceeds from rest at infinite
initial separation,  which implies that the conserved energy
$E(z)= E(\infty)=0$.
In case (B),
the radial infall proceeds from rest at finite
initial separation $z_{0}$, which implies
\begin{equation}
E(z = z_{0}) = -\mu c^2 \gamma_{0} \left\{ 1 -{1 \over 2}\,\gamma_{0}
+{1 \over2}\left[ 1
+\frac{15}{2}\,\eta\right]\,\gamma_{0}^{2}\right \}\,,
\label{rfbe}
\end{equation}
where $\gamma_0= {G\,m/z_0 c^2}$.
Inverting $ E(z) $ for $ {\dot z}^2 $ and using Eq. (\ref{rfbe})
we obtain
\begin{eqnarray}
\dot{z} &=& -c\left\{ 2\left(\gamma -\gamma_{0}\right)\left[ 1
-5{\gamma}\left(1 -\frac{\eta}{2}\right)
+{\gamma_{0}}\left(1
-\frac{9\eta}{2}\right)\right.\right.\nonumber\\
& &\left.\left.+{\gamma^{2}}\left( 13 -\frac{81\eta}{4}
+5\eta^{2}\right)
-{\gamma\gamma_{0}}
\left( 5 -\frac{173\eta}{4} +13\eta^2\right)
\right.\right.\nonumber\\
& &\left.\left. +{\gamma_{0}^{2}}
\left( 1 -\frac{5\eta}{4} + 8\eta^2\right)
\right]\right\}^{1\over2}\,,
\label{rf3}
\end{eqnarray}
where $\gamma= G\,m/z\,c^2$. Using the radial infall 
restrictions and Eq. (\ref{rf3}) in Eqs.
(\ref{gief})
we obtain for the case B),the far-zone radiative energy flux 
\begin{eqnarray}
\left( {d{\cal E} \over dt}\right)_{\rm far-zone}^{\rm inst} 
&=& { 16 \over 15}\,c^5\,\eta^2\,\gamma^5
\left \{ 1 - x -{ 1\over 7}\left [ 43 -{111 \over 2}\eta
-x( 116 -131\eta ) + x^2( 71 -{135\,\eta \over 2})
\right ]{\gamma}\right.\nonumber \\
& & \left.
- {1 \over 3} \left [ { 1127 \over 9} + {803 \over 12}\eta
-112\,\eta^2
+ { x \over 7}\left( { 4471 \over 9} - {15481 \eta \over 3}
+ 2864\,\eta^2 \right )\right.\right.\nonumber \\
& & \left.\left.
- { x^2 \over 7}\left ( 1870
-{ 38521\,\eta \over 6} + { 8800\,\eta^2 \over 3} \right )
+x^3 \left ( 83 -{1183\,\eta \over 4} + {872\,\eta^2 \over 7}\right )
\right ]
{\gamma^2} \right \}\,,
\label{refA}
\end{eqnarray}
where $ x = {\gamma_0 / \gamma} $.
For the case (A), the expressions for
$ \dot z $ and $ { d{\cal E}/ dt } $ are obtained by
setting $ \gamma_0 =0 $ in Eqs. (\ref{rf3}) and ( \ref{refA}).
Eq. ( \ref{refA}), along
with corresponding one  for case (A) are in agreement
with \cite{SPW96}.
\section{The evolution of the orbital elements in the generalized 
quasi-Keplerian parameterization of the binary} 
\label{sec:dpdt2}

 In this section, we compute
the 2PN corrections  to the rate of decay of the orbital elements of  
a compact binary, in quasi-elliptical orbit,
{\it i.e }
the effect of the 4.5PN radiation  reaction on
a 2PN accurate conservative elliptical motion, extending the earlier
computations\cite{BS89,BS93,JS92}.
The basic ingredients we employ for the calculations are the far-zone 
energy and angular momentum fluxes in the harmonic coordinates computed 
in previous sections
and a 2PN accurate description of  the relative
motion of the compact binaries available in a generalized quasi-Keplerian
parameterization given in the ADM coordinates \cite{DS88,SW93,NW95}.
Since the De Donder(harmonic) and the ADM  coordinates are different
at the 2PN order,
we use  the coordinate transformations connecting the 
harmonic and the ADM coordinates\cite{DS85}, to rewrite the far-zone fluxes 
in the ADM coordinates. The far zone fluxes, in the ADM coordinates
are averaged over an orbital period,
extending the earlier computations  at the 1PN and the 1.5PN order
\cite{BS89,BS93,JS92,RS97}.
The 2PN corrections to the rate of decay of the orbital elements are
computed using heuristic arguments based on the conservation 
of energy and angular momentum
to the 2PN order.
Before proceeding to the actual computations, in the next two sections,
we summarize the generalized quasi-Keplerian description of
the binary orbits in the ADM coordinates and the transformations needed,
to relate the kinematical variables in the harmonic 
and the ADM coordinates.  
\subsection{ The second post- Newtonian motion of compact binaries }

The generalized quasi-Keplerian description 
for the general binary orbits to the 2PN order,
developed by  Damour, Sch\"afer, and Wex
\cite{DS88,SW93,NW95} is best suited for the calculation
we propose to do in the following sections and we summarize it 
in what follows. 
Let $r_A(t_A),\, \phi_{A}(t_A)$ be the planar 
relative motion of the two compact objects
in usual polar coordinates associated with the ADM coordinates.
The radial motion $r_A(t_A)$  is conveniently  parameterized  by
\begin{mathletters}
\begin{eqnarray}
r_A &=& a_r \left ( 1 -e_r \cos u\right )\,,\\
n( t_A -t_0 ) &=& u -e_t \,\sin u + { f_t \over c^4}\sin v
+ { g_t \over c^4}\left ( v -u\right )\,,
\end{eqnarray}
\label{admrn}
\end{mathletters}
where `$u$' is the `eccentric anomaly' parameterizing the motion and the
constants $a_r,\, e_r,\, e_t,\, n$ and $t_0$ are 
some 2PN semi-major axis, 
radial eccentricity,  time eccentricity,  mean motion, and 
initial instant respectively. The angular motion $\phi_A(t_A)$ 
is given by
\begin{mathletters}
\begin{eqnarray}
\phi_A -\phi_0 &=& \left ( 1 +{ k \over c^2} \right )v
+ { f_{\phi} \over c^4} \sin 2v + { g_{\phi} \over c^4}\sin 3v\,,\\
\mbox{ where } v &=& 2\, \tan^{-1}\left \{ \left ( 
{1 +e_{\phi} \over {1 -e_{\phi}}}
\right)^{ 1\over 2} \, \tan ( { u \over 2}) \right \}\,.
\end{eqnarray}
\label{admphi}
\end{mathletters}
In the above~  $\phi_0, k, e_\phi$ are some constant,  periastron
precession constant, and  angular eccentricity respectively. 
All the parameters $n,\, k,\, a_r,\, e_t,\, e_r,\, e_{\phi},\, 
f_t,\, g_t,\, f_{\phi} $ and $g_\phi$ are
functions of the 2PN conserved  energy and angular momentum
per unit  reduced mass\,$\mu$: 
To avoid additional notations following 
\cite{DS88,SW93,NW95}, these are also denoted as  
$ E $ and $|{\bf J}|$. 
Their
explicit functional forms, given in \cite{SW93}
are displayed below
\begin{mathletters}
\begin{eqnarray}
a_r &=& -{G\,m \over 2\,E}\left \{ 1 + { 1\over 2\,c^2}(7 -\eta)\,E
+ {1 \over c^4}\left [ { 1\over 4}( 1 +10\eta +\eta^2)E^2 
+{1 \over 2}
(17 -11\eta)\,{E \over h^2} \right ]\right \} \,,\\
e_r^2 &=& 1+2Eh^2 -{ 1\over c^2}\biggl \{{2}(6 -\eta)E +{5}( 3
-\eta)\,E^2h^2 \biggr \}
+{1 \over c^4}\biggl \{ (26 +\eta +\eta^2)E^2 \nonumber \\
 & & -{2}( 17 -11\eta)\,{E \over h^2} +
( 80 -55\eta +4\eta^2)E^3\,h^2 \biggr \} \,,\\
n &=& {( -2\,E)^{3 \over 2} \over Gm } \biggl \{ 1  
+ { 1 \over 4\,c^2}( 15 -\eta)E + { 1 \over c^4}
\biggl [{ 1\over 32}(555 +30\eta
+11 \eta)E^2 -{ 3\over 2}
(5 -2\eta)\,{(-2E)^{3 \over 2} \over h} \biggr ]\biggr \}\,,\\
e_t^2 &=& 1 +2Eh^2 + {1 \over c^2}\biggl \{ 4( 1 -\eta )E + 
( 17 -7\eta)E^2\,h^2 \biggr \} 
+{ 1 \over c^4}\biggl \{ 2( 2 +\eta +5\eta^2) E^2 
\nonumber \\
&& - ( 17 -11\eta)\,{E \over h^2} 
+ ( 112 - 47\eta + 16\eta^2 )E^3h^2 -
{3}( 5 -2\eta)(1 +2Eh^2){(-2E)^{3 \over 2} \over h}\biggr \}\,,\\
f_t &=& -{ 1 \over 8\, h}\, \eta ( 4 + \eta)
{( 1 +2 E h^2 )}^{1 \over 2}(-2 E)^{3 \over 2}\,,\\
g_t &=& {3\over 2}\,(5 -2\eta) {(-2E)^{3 \over 2} \over h}\,,\\
k &=& { 3 \over h^2}\biggl \{ 1 + { 1\over 2\,c^2 }\biggl [( 5 -2\eta)E +
{ 5 \over 2\,h^2}( 7 -2\eta)\biggr ] \biggr \} \,,\\
f_{\phi} &=& { 1 \over 8}\,{\eta \over h^4}\,( 1 -3\eta)( 1 +2E\,h^2)\,,\\
g_{\phi} &=& -{ 3 \over 32}\,{\eta^2 \over h^4}
( 1 +2\,Eh^2)^{ 3/ 2}
\,,\\
e_{\phi}^2 &=&  1 +2Eh^2 -{ 1 \over c^2 }\biggl \{ 12 E 
+ ( 15 -\eta )E^2\,h^2 \biggr \}
-{ 1 \over 8\,c^4} \biggl \{ 4( 16 -88\eta -9\eta^2 )E^2
\nonumber \\
& & 
- 4( 160 -30\eta +3\eta^2)E^3h^2  +
( 408 -232\eta -15\eta^2){ E \over h^2} \biggr \}\,,
\end{eqnarray}
\label{nwexp}
\end{mathletters}
where $h={|{\bf J}|/( G\,m)}$.
Note that $ n = { 2\pi / P}$, where $P$ is the period of
the binary.
Using these parametric
equations of the motion, we compute 
$\dot{r}^{2}_{A},\,v^{2}_{A}$ to the 2PN order   
in terms of $ E, h^2, (1 -e_r\cos u) $  using, 
\begin{mathletters}
\begin{eqnarray}
\frac{dt_A}{du}&=&\frac{\partial t_A}{\partial u}+
\frac{\partial t_A}{\partial v}\frac{dv}{du}\\
\dot r^{2}_{A} &=&\left ({{ dr_{A}\over du }/ {d t_A\over du}}\right )^2\\
\dot {\phi}^{2}_{A} &=& \left ( {{ d\phi_{A} \over dv}{d v\over du}
 / { d t_A\over du}} \right )^2\\
v^{2}_{A} &=& \dot r^{2}_{A} + r^{2}_{A}\dot { \phi}^{2}_{A}\,.
\end{eqnarray}
\label{ffbaa}
\end{mathletters}
The  subscript `${\rm A}$ ' present in  Eqs. (\ref{ffbaa}) is a reminder 
that the expressions refer to the  ADM gauge. We have  
\begin{mathletters}
\begin{eqnarray}
\dot r^2_{\rm A} &=&  \left \{ -1 + {2 \over (1 -e_r\cos u)} 
+{2 \over (1 -e_r\cos u)^2}\,E\,h^2 \right \}(-2E)
\nonumber \\
& & + { 1 \over c^2}\biggl \{ -3 +9\eta  +{ 1 \over (1 -e_r\cos u)} 
\biggl [ 38 -30\eta \biggr ]
\nonumber \\
& &
-{ 1 \over (1 -e_r\cos u)^2} \biggl [ 40  -20\eta 
- \left ( 36 -28\eta \right )\,Eh^2 \biggr ]
 -{ 1 \over (1 -e_r\cos u)^3} \biggl [ \left (64 -24\eta \right )Eh^2 \biggr ]
\biggr \} E^2
\nonumber \\
& & - { 1 \over c^4}\biggl \{ 4 -19\eta +16\eta^2 
-{1 \over (1 -e_r\cos u)}\biggl [ 168 -326 \eta +98\eta^2
- {1 \over E\,h^2}\left ( 34 -22 \eta \right )
\biggr ]
\nonumber \\
& &
+{1 \over (1 -e_r\cos u)^2}\biggl [ 496 -712 \eta + 164\eta^2
-\left (213 -298  \eta + 85\eta^2 \right )\,Eh^2\biggr ] 
\nonumber \\
& &
-{1 \over (1 -e_r\cos u)^3}\biggl [212 -332\eta +80\eta^2
-\left ( 800 -932\eta +188\eta^2 \right )\,Eh^2\biggr ]
\nonumber \\
& &
-{ 1 \over (1 -e_r\cos u)^4}\left [ 528 -528\eta +96 \eta^2 \right ]\,Eh^2
+ { 1\over (1 -e_r\cos u)^5}\left [ 32 +8\eta^2 \right ]\eta\,E^2h^4
\biggr \} (-E)^3 \,,
\nonumber \\
\\
v^2_{A} &=& \biggl \{ -1 + { 2 \over (1 -e_r\cos u)} \biggr \}(-2E)
\nonumber \\
& &
-{ 1 \over c^2}\biggl \{ 3 -9\eta -{1 \over (1 -e_r\cos u)}\left [ 38 -30\eta \right ]
+{1 \over (1 -e_r\cos u)^2}\left [ 40 -20\eta \right ] 
\nonumber \\
& &
+ 8\,{1 \over (1 -e_r\cos u)^3}\eta\,Eh^2
\biggr \}E^2
\nonumber \\
& &
-{ 1\over c^4}\biggl \{ 4 -19\eta +16\eta^2 -{ 1 \over (1 -e_r\cos u)}\biggl [
168 -326\eta +98\eta^2 - 
{1 \over Eh^2}\left ( 34 -22 \eta \right )\biggr ] 
\nonumber \\
& &
+{ 1 \over (1 -e_r\cos u)^2}\biggl [ 428 -668\eta +164\eta^2 \biggr ]
\nonumber \\
& &
-{ 1 \over (1 -e_r\cos u)^3}\biggl [ 212 -332 \eta +80 \eta^2 -\left ( 76 
-84 \eta \right )\eta\, Eh^2 \biggr ]
\nonumber \\
& &
-{ 1\over (1 -e_r\cos u)^4}\left [ 80 - 128 \eta \right ]\eta \,Eh^2
+ 72\,{ 1 \over (1 -e_r\cos u)^5}\,\eta^2 \,E^2h^4  
\biggr \}(-E)^3\,.
\end{eqnarray}
\end{mathletters}
These expressions for 
$ {\dot r}^2_{A} $ and $ v^2_{A}$
are consistent with Eqs. (6) and (7) of \cite{WR95}.

\subsection{ The  transformation between  
De-Donder ( harmonic ) and ADM gauges}

As pointed out earlier,  the  far-zone fluxes obtained 
in previous sections are in the harmonic coordinates,
whereas, the 2PN accurate orbital description given by Eqs. (\ref {admrn}),  
(\ref{admphi}), and (\ref{nwexp})  are in the ADM coordinates.
For the purpose of averaging the far zone fluxes using the 
the 2PN accurate orbital representation, 
we need to go 
from the  
De-Donder(harmonic) to the ADM gauge, 
 and rewrite the  expressions for the far-zone fluxes  in the
ADM coordinates. These follow straightforwardly from the transformation
equations in \cite{DS85} and we list below the   
transformation equations,  relating the  harmonic(De-Donder) variables to the
corresponding ADM variables: 
\begin{mathletters}
\label{ctad}
\begin{eqnarray}
{\bf r}_{ \rm D} &=& {\bf r}_{ \rm A}
+ { Gm \over 8\,c^4\,r}\left \{ \left [ \left ( 5 v^2
-\dot r^2 \right )\eta +  2 \,\left ( 1
+12\eta \right ){ Gm\over r} \right ] { \bf r}
\right.\nonumber\\
& & \left.
-  18 \,\eta \, r\dot r\,{\bf v} \right \}\,,\\
t_{\rm D}  &=& t_{ \rm A } - { Gm \over c^4}\,\eta \,\dot r \,,\\
{ \bf v }_{ \rm D} &=& { \bf v }_{ \rm A} - { Gm \dot r\over 8\,c^4\, r^2}
\biggl \{ \biggl [ 7 v^2 + 38 { Gm \over r}
- 3\dot r^2 \biggr ]\eta + 4\,{ G m \over r}\, \dot r \biggr \}
{ \bf r}
\nonumber \\
& &
- { Gm \over c^4 r} \biggl \{ 
\biggl [ 5 v^2 - 9 \dot r^2 - 34 { G m\over r} \biggr ]\eta
- 2\,{ Gm \over r} \biggr \}{ \bf v} \,,\\
({\bf L}_{N})_{\rm D}&=& ({\bf L}_{N})_{\rm A}\biggl \{
1 + { G\,m \over 4\,c^4\,r}\biggl [ ( 2 +29\eta ){G m \over r}
+ 4\,\eta \,\dot r^2 \biggr ]\biggr \} \,,\\ 
r _{ \rm D} &=& r _{ \rm A} + { G m \over 8\,c^4}
\left \{  5\, \eta v^2 + 2\,
 \left ( 1 +12\eta \right ){ G m\over r} -  19\, \eta \dot r^2
\right \}\,,\\
v^2_{ \rm D} &=& v^2_{\rm A} -
{ G m\over 4\,c^4\,r}\biggl \{ \left [ 5 v^4 -2v^2\dot r^2
-3\dot r^4 \right ]\eta
\nonumber\\
& & 
-\biggl [ 2\left ( 1 +17\eta \right )v^2
-\left ( 4 +38\eta \right )\dot r^2 
\biggr ]{ Gm \over r}\,\biggr \}\,,\\
{\dot r^2}_{\rm D} &=& {\dot r^2}_{\rm A}
- { Gm \over 2\,c^4\,r}\,\dot r^2 \left \{ 15\left ( v^2 -\dot r^2 \right )\eta
+ \left ( 1 +2\eta \right ){ G m\over r} \right \}\,.
\end{eqnarray}
\end{mathletters}

The subscript `${\rm D}$' denotes  
quantities in the De-Donder~( harmonic ) coordinates.
Note that in all the above  equations the differences between the
two gauges are of the 2PN order.
As there is no difference between the harmonic and the ADM coordinates
to 1PN  accuracy, in Eqs. (\ref{ctad}), for the 2PN terms no suffix  
is  used. 
The 2PN extension  of the evolution of
the orbital elements  thus requires more technical care than the 1PN
case due to the differences in the ADM and 
harmonic coordinates given by Eqs. (\ref{ctad}).
Finally using the above equations we have verified that the 
expressions given by Eqs. (\ref{ddcom}), relating the individual
locations of the two bodies to the centre of mass coordinate are
consistent with the corresponding choice in ADM coordinates, given
by Eqs. (A5) - (A8) of \cite{NW95}.

\subsection { 2PN corrections  to $ <{ d {\cal E} / dt}>$ and 
$ <{d {\cal J} /dt }>$ }

Starting from Eqs.(\ref{gief}) and (\ref{giamf}) for the far-zone
fluxes in the harmonic coordinates 
we use   Eqs. ( \ref{ctad}), 
to obtain  ${ d{\cal E} / dt } $ and $ { d{\bf {\cal J}} / dt } $
~in the ADM coordinates.
For economy of presentation, we write the results  in the following manner,
$ (Flux)_{\rm A} = (Flux)_{\rm O} + $ `$ Corrections $', where  
$(Flux)_{\rm A}$ represent 
the far-zone flux in the ADM coordinates.
$(Flux)_{\rm O}$ is a short hand notation for 
expressions on the r.h.s of Eqs. (\ref{gief}) and (\ref{giamf}),
where $ v^2, \dot r, r $ are the ADM variables 
$v_{\rm A}^2, {\dot r}_{\rm A}, r_{\rm A} $ respectively.  
For example, the Newtonian part of $({d {\cal E}/ dt})_{\rm O}$ 
will be  $ {8 \over 15}\frac{G^3\,m^2\,\mu^2}{c^5\,r_{\rm A}^4}
\left\{ 12v_{\rm A}^2 - 11{\dot r}_{\rm A}^2 \right\} $.
The `$ Corrections $' represent the differences at the 2PN 
order, that arise due to the change of the coordinate system,
given by Eqs. (\ref{ctad}). As the two coordinates are
different at the 2PN order, the `$ Corrections $' come 
only from the leading Newtonian terms in Eqs. (\ref{gief}) and (\ref{giamf}). 
\begin{mathletters}
\label{admfxs}
\begin{eqnarray}
\biggl (\frac{d{\cal E}}{dt}\biggr )_{\rm A}&=&
\biggl (\frac{d{\cal E}}{dt}\biggr )_{\rm O}
- { G^4m^3\mu^2 \over 15 c^9 r_{\rm A}^5}\biggl \{
\left [ ( 48 +336\eta)v_{\rm A}^2 -(36 +232\eta)\dot r_{\rm A}^2 
\right]{Gm \over r_{\rm A}}
\nonumber \\ & &
+  \left[ 360 v_{\rm A}^4 -1840 v_{\rm A}^2\dot r_{\rm A}^2 +
1424 \dot r_{\rm A}^4\right] \eta \biggr \}
\\
\biggl (\frac{d{\bf { {\cal J} }}}{dt}\biggr )_{\rm A}&=&
\biggl (\frac{d{\bf {\cal J}}}{dt}\biggr )_{\rm O}
+ { G^3m^2\mu^2 ({\bf {\tilde L}}_N)_{\rm A} \over 5\, c^9 r_{\rm A}^4}\biggl \{
\left [ ( 4 +68\,\eta)v_{\rm A}^2 
-(8 +76\eta){Gm \over r_{\rm A}} +(2 +82\eta)\dot r_{\rm A}^2
\right ] {Gm \over r_{\rm A}} \nonumber \\
& &+ (363  v_{\rm A}^2\dot r_{\rm A}^2 
- 50 v_{\rm A}^4 - 363 \dot r_{\rm A}^4) \,\eta \biggr \}
\end{eqnarray}
\end{mathletters}
Note that all the variables on the r.h.s of Eqs. (\ref{admfxs})
are in the ADM coordinates. 
In the circular limit energy and angular momentum fluxes are again
related as in Eqs. (\ref{hcirl}), via the corresponding 
`$v^2$' in the ADM coordinates  given by
\begin{equation}
v_{\rm A}^2 ={Gm\over r_{\rm A}}\left \{ 1 -(3 -\eta){Gm \over
c^2\,r_{\rm A}} 
+{1 \over 8} (42 -5\eta + 8\eta^2){ G^2m^2 \over c^4\,r_{\rm A}^2}
\right \}\,.
\end{equation}

From this point onwards, in this section, we  work 
exclusively in  the ADM gauge  and hence we drop the subscript `{\rm A}' 
for the ease of presentation. 
We now have all the ingredients needed to   calculate the  2PN 
corrections in  $<d{\cal E}/dt>$ and $<d{ {\cal J}}/dt>$. We explain in detail, 
the procedure to compute $<d{\cal E}/dt>$ and only display the 
final expression for $<d{\cal J}/dt>$, as  the  procedure  
is the same in both the 
cases. 
Starting from  Eqs. (\ref{admfxs}),
(\ref{gief}), and (\ref{giamf}) 
which give the far zone fluxes as functions of $v^2\,,\, \dot{r}^2\,,\,$
and ${ G\,m/r}$,  
 we use the 2PN accurate orbital representation, to rewrite 
$d{\cal E}/dt$ as a polynomial in 
$(1-e_r\cos u)^{-1}$. This
polynomial is of the form
\begin{equation}
{d{\cal E} \over dt} = \frac{du}{ndt}\sum_{N=2}^{8}
\frac{\alpha_N(E, h)}{(1- e_r\cos u)^{(N+1)}}\,,
\label{bs413}
\end{equation}
where for the convenience we have factored out ${du}/{ndt}$ given by
\begin{eqnarray}
{ du \over ndt} &=& { 1 \over (1 -e_r \cos u)}\biggl \{ 1 - {E \over c^2}\,( 8 -3\eta)
\biggl ( 1 -{1 \over (1 -e_r \cos u)} \biggr ) 
\nonumber \\
& &
+ {1 \over 2\,c^4}\biggl [ 
E^2\biggl ( ( 56 -63\eta +6\eta^2)  \nonumber \\
& & -{1 \over E\,h^2}\,(17 -11\eta)(1 -{1 \over (1 -e_r \cos u)})
-{ 1 \over (1 -e_r \cos u)}( 184 -159\eta +24\eta^2) 
\nonumber \\
& &
+{ 1 \over (1 -e_r \cos u)^2}( 68 -76\eta +17\eta^2) 
-{ 2\,E\,h^2 \over (1 -e_r \cos u)^3}\,\eta ( 4 +\eta) \biggr )
\nonumber \\
& &
+ {3 \over h}\,(-2\,E)^{3/2}(5 -2\eta) \biggr ]\biggr \} \,.
\end{eqnarray}
It is
a straightforward algebra to show that the coefficients $\alpha_N(E,h)$ in
Eq. (\ref{bs413}) take the form
\begin{equation}
\label{bs414}
\alpha_{ \rm N}( E,h) = { \eta^2 \over G\, c^5}(-E)^5 \beta_{ \rm N}(E,h)\,,
\end{equation}
where $\beta_{ \rm N}(E,h)$ for  $ \rm N =1,2, \ldots 8$ are given by
\begin{mathletters}
\label{bs415}
\begin{eqnarray}
\beta_2 &=& -{256 \over 15} + { 1 \over 105\, c^2}(298\,24 - 154\,88\eta)E
+{ 1 \over c^4}\biggl \{- { 1 \over 315}( 791\,168 - 874~624 + 179456\eta^2)E^2
\nonumber \\
& & 
+ { 128 \over 5}(17 -11\eta){ E \over h^2}
+ {1 \over 5} ( 640 -256\eta) 
 { (-2E)^{3 \over 2} \over h}\biggr \}\,,\\
\beta_3 & = & { 512\over 15} -{ 1\over 35\,c^2}( 263~68 -19968\eta )E
 +{ 1 \over c^4} \biggl \{ \biggl [ { 2716928 \over 315}
- { 13040896 \over 945}\eta
+ { 538496  \over 135}\eta^2 \biggr ]E^2   \nonumber\\
&& 
-{ 896 \over 15}(17 -11\eta )
{ E \over h^2}
-{1 \over 5} ( 1280 -512\eta ) 
{ (-2E)^{3 \over 2} \over h}\biggr \}\,, \\
\beta_4 &=& -{5632 \over 15}Eh^2
+ { 1 \over c^2}\left \{ { 1 \over 7}( 1024 -3072\eta)E
+{ 512 \over 105}( 1729 -930\eta)\,E^2\,h^2 \right \} \nonumber \\
& &+ { 1\over c^4} \left \{( {46840064 \over 2835} + { 3537664 \over 945}\eta
- { 2315648 \over 315}\eta^2 )E^2 \right. \nonumber\\
& &\left.
-{ 128 \over 105}( 86403 -89968\eta
+20923\eta^2)\,E^3h^2
 -{ 256 \over 15}(17 -11\eta){ E \over h^2}
\right. \nonumber\\
& & \left.
- { 1\over 5}( 7040 -2816\eta){ (-2E)^{ 5\over 2}h}  \right \}\,, \\
\beta_5 &=& -{512 \over 105 \,c^2}( 3232 -1395 \eta)\, E^2\,h^2
+ { 1\over c^4}\biggl \{ -\biggl [ { 14200576 \over 2835} 
 -{ 38656 \over 189}\eta
- { 219904 \over 63}\eta^2 \biggr ]\,E^2  \nonumber\\
& & 
+ { 256 \over 945}\biggl [ 148\,648\,8 -1545569\eta
+ 343813 \eta^2 \biggr ]\, E^3h^2 \biggr \} \,,\\
\beta_6 &=& -{ 512 \over 35 \,c^2}( 687 -620 \eta) E^3h^4
-{ 1 \over c^4} \biggl \{ { 256 \over 945}
\biggl [1221526 -1333624 \eta
+ 319739 \eta^2 \biggr ]\,E^3h^2
\nonumber \\
& &-{ 512 \over 105}\biggl [ 51396
- 91541\eta +27508\eta^2 \biggl ]\,E^4h^4  \biggr \} \,, \\
\beta_7 &=& -{512 \over 945\,c^4}\biggl \{ 748\,032 - 1385005 \eta
+ 387911\eta^2\biggr \}\,E^4h^4\,,\\
\beta_8 &=& -{4096 \over 315\,c^4}\biggl \{ 2501 -202~34 \eta 
+8404 \eta^2 \biggr \}\,E^5h^6\,.
\end{eqnarray}
\end{mathletters}
To the 1PN order Eqs. (\ref{bs415}) agree with Eqs.~(4.15) of \cite{BS89}.
 The far-zone energy flux $(d{\cal E}/dt)$ is  a periodic function 
of time with period $P =2\pi/n$.
Averaging $(d{\cal E}/dt)$, given by Eqs. (\ref{bs413}), (\ref{bs414})
and (\ref{bs415})  
over one time period P, we obtain
\begin{equation}
< { d{\cal E} \over dt }> 
={1 \over P}\int^{P}_{0}\,{d {\cal E} \over dt }\biggl (t \biggr )\,dt\,
={ 1 \over 2\,\pi}\int ^{2\,\pi}_{0}\biggl 
({ndt\over du} \biggr ){d {\cal E} \over dt }\biggl (u \biggr ) du \,.
\label{bs417}
\end{equation}
The integrals in Eq.(\ref{bs417})  
are the Laplace second integrals for
the Legendre polynomials \cite {WW1927} which yield,
\begin{eqnarray}
{1 \over 2\pi}\int ^{2\pi}_{0} { du \over \biggl \{ 
1 -e_r\cos u\biggr \}^{N+1}}
&=& {1 \over ( 1-e_r^2)^{ N+1 \over 2}}P_{\rm N} \biggl(
 { 1\over \sqrt {(1 -e_r^2)}}\biggr ),
\label{bs419}
\end{eqnarray}
where $P_{ \rm N }$ is  Legendre polynomial. 
Using Eq. ( \ref{bs419})
in Eq. (\ref{bs417}), we obtain $<d{\cal E}/dt>$ in terms of $ E$ and $e_r$:
\begin{eqnarray}
< {d {\cal E} \over dt }> &=& { 1024 \over 5}
{\mu \,\eta  \over  G\,m\, c^5 }\,{ (-E)^5 \over 
( 1 -e_r^2)^{ 7 \over 2}}
\biggl \{ 1 + { 73 \over 24}e_r^2 + { 37 \over 96}e_r^4
\nonumber \\
& & +{ 1 \over 168}\, {(-E) \over c^2 \,(1 -e_r^2)}
\biggl  [ 13 - 6414 e_r^2
- { 274\,05 \over 4}e_r^4 - {537\,7 \over 16}e_r^6
 \nonumber \\
& &  -
\left ( 840 + { 6419 \over 2}e_r^2 + { 5103 \over 8}e_r^4
- {259 \over 8}e_r^6 \right )\,\eta
\biggr ] \nonumber \\
& & - {(-E)^2  \over c^4} \biggl [ 
{ 1 \over 16\,(1 -e_r^2)^{5 \over 2}}
\biggl ( (480 -192\eta) + ( 500 -200\eta)e_r^2 - ( 2255 -902 \eta)e_r^4
\nonumber \\
& & 
+ ( 1090 -436 \eta)e_r^6 + ( 185 -74 \eta)e_r^8 \biggr )
 \nonumber \\
& & - { 1  \over (1 -e_r^2)^2 } \biggl ( { 253\,937 \over 4536}
- { 180\,65 \over 504}\eta +10 \eta^2 - \left ( { 879\,749 \over 4536 }
- { 301\,37 \over 72}\eta - { 1877 \over 48 }\eta^2 \right )e_r^2
 \nonumber \\
& & - \left ( { 513\,337 \over 6048 } - { 531\,871 \over 672}\eta
+ { 1139 \over 192}\eta^{2} \right ) e_r^4
 \nonumber \\
& &  + \left ( { 249\,479\,5 \over 8064}
+ { 4823 \over 128 }\eta - { 383 \over 96}\eta^2 \right )e_r^6
+ \left ( { 283\,685 \over 16128} - { 131\,47 \over 2688}\eta
+ { 37 \over 192}\,\eta^2 \right ) e_r^8
\biggr )
\biggr ]
\biggr \}
\label{bs421}
\end{eqnarray}
Following exactly a similar procedure, we obtain the 2PN correction to
$<d{ {\cal J}}/dt>$.  The final result we obtain is: 
\begin{eqnarray}
<{ d {\cal J} \over d t}> &=& { 4  \over 5}\, {\mu\,\eta \over c^5}
{(-2 E)^{7 \over 2} \over  ( 1 -e_r^2)^3 }
\biggl \{ 8 - e_r^2 - 7e_r^4
 \nonumber \\
& &
-{ (-E) \over 168 \,c^2} \biggl [ \left ( 292\,0 + 705\,6 \eta \right )
+ \left ( 197\,38 + 144\,34 \, \eta \right ) e_r^2
+ \left ( 127 + 133\,0\,\eta \right )e_r^4 \biggr ]
\nonumber \\
& &
- { (-E)^2 \over c^4} \biggl [ { 1 \over (1 -e_r^2)^{1 \over 2}}
\biggl ( 240 -96\eta - ( 30 - 12\eta)e_r^2 
-( 210 -84\eta)e_r^4 \biggr )
\nonumber \\
& & 
- { 1 \over (1 -e_r^2)} \biggl ( { 299\,623 \over 1134}
- { 220\,25 \over 252}\,\eta + { 351 \over 4}\,\eta^2
\nonumber\\
& & 
- \left ( { 131\,627\,3 \over 864 } - { 815\,597 \over 336 }\,\eta
- { 292\,07 \over 96}\,\eta^2 \right )e_r^2
\nonumber \\
& &
- \left ( { 290\,113\,3 \over 6048 } - { 124\,403 \over 96}\,\eta
- { 718\,7\,\over 48 }\eta^2 \right )e_r^4
+ \left ( { 7526 \over 63} - {2869 \over 224 }\eta
+ { 143\,5 \over 96 }\,\eta^2 \right )e_r^6
\biggr )
\biggr ] \biggr \}
\label{js421}
\end{eqnarray}
To the 1PN order,  Eqs. (\ref{bs421}) and (\ref{js421})
 agree with \cite{BS89,JS92} as required.
For the special case of circular orbits, $ e_r=0$ and we observe that, 
$ <{ d {\cal E} / d t}> =\omega \,<{ d {\cal J} / d t}>$ to the 2PN
order, where $\omega $, the mean angular frequency of the 
relative motion, defined by $\omega = n(1 +k )$ is given by
\begin{equation}
\omega = { (-2\,E)^{3 \over 2} \over G\,m}\left \{
1 - { 1\over 4\,c^2}(9 +\eta)E +
{ 1\over 32\,c^4}(2811 -1170\eta +11\eta^2)E^2\right \}
\end{equation}
   
                      It is not very difficult to trace 
the origin of the two types of terms
in Eqs. (\ref{bs421}) and (\ref{js421}) at the 2PN order.
It is related to the fact that `$ Corrections$' in Eqs.
(\ref{admfxs}), arising from the transformation equations
connecting the harmonic and the ADM coordinates have a different functional
form than the 2PN contributions to the corresponding far-zone
fluxes in the harmonic coordinates. For example, in the  far-zone 
energy flux, `$ Corrections$' contain a common factor
$({G^4\,m^3/r^5})$, unlike the 2PN contributions in harmonic 
coordinates which have only 
$({G^3\,m^2/r^4})$ as the common factor ({\it c.f } Eqs.
({\ref{gief}}) and ({\ref{admfxs}})). These  different functional
forms, after the  averaging procedure  give  rise to the two 
different types of terms in Eqs. (\ref{bs421}) and (\ref{js421}).
           
	   We display below  $<{ d {\cal E} /d t}>$ 
and $<{ d {\cal J} / d t}>$
in terms of $ { G\,m/ a_r} $ and $ e_r $, which can easily
be obtained from Eqs. (\ref{bs421}) and (\ref{js421}),
using $E$ written in terms of $ { G\,m/a_r} $ and $ e_r $ to 
the 2PN order. 
The required equation for E is obtained  from
Eqs.(\ref{nwexp})  for $a_r$ and $e_r$ 
by inverting them for $E$ and $h^2$ respectively order by order.
Eliminating $h^2$ from the expression for $E$  we finally get, 
\begin{eqnarray}
E &=& -{ c^2 \over 2}\,\zeta \biggl \{ 1 - {1 \over 4} (7 -\eta)
\zeta 
+ { 1 \over 8} \left [ ( 25 -2\eta +\eta^2) - 2\,{(17 -11\eta) \over
 ( 1 -e_r^2) } \right ]\,\zeta^2 \biggr \},
\end{eqnarray}
where $\zeta ={ G\,m/{c^2a_r}}$.
Using the above expression for $E$, Eq. (\ref{bs421}) becomes 
\begin{eqnarray}
\label{js932}
<{ d {\cal E} \over d t}>&=&{ 1 \over 15}\, {c^5\over G}\,\eta^2
{\zeta^5 \over (1-e_r^2)^{13 \over 2}}
\biggl \{ \biggl [ (96 +292 e_r^2 +37 e_r^4) ( 1 -e_r^2)^3 
\biggr ]\nonumber \\
&& -{ 1\over 56} {\zeta } (1 -e_r^2)^2
\biggl [ ( 468\,32 + 672\,0\eta) + ( 198\,664 +376\,32\eta)e_r^2
\nonumber \\
& &
-( 153\,30 -280\,56\eta)e_r^4 -( 127\,53 -207\,2\eta)e_r^6 
\biggr ]
\nonumber \\
& & +{ \zeta^2 } \biggl [ { 1 \over 6048}( 1 -e_r^2)
\biggl ( ( 224\,053\,12 + 122\,492\,16 \eta) 
\nonumber \\
& &
+ ( 912\,416\,00 + 973\,409\,76\eta + 290\,304 \eta^2 )e_r^2
\nonumber \\
& &
- ( 977\,677\,44 - 731\,619\,00 \eta -239\,500\,8 \eta^2)e_r^4
\nonumber \\
& &
- ( 757\,105\,2 + 606\,592\,8 \eta - 280\,627\,2\eta^2 )e_r^6
\nonumber \\
& &
+( 680\,528\,7 -148\,921\,2\eta + 223\,776\eta^2 )e_r^8
\biggr ) \nonumber \\
& & - { 3\over 2} ( 1-e_r^2)^{5 \over 2} 
\biggl ( ( 96 +292 e_r^2 +37e_r^4)( 5 -2\eta)\biggr )
\biggr ]
\biggr  \}
\end{eqnarray}
while Eq. (\ref{js421}) gets transformed to,
\begin{eqnarray}
\label{js929}
<{ d {\cal J} \over d t}>&=& {4 \over 5}{ \mu\,\eta  c^2}
{\zeta^{7 \over 2}\over (1 -e_r^2)^4}
\biggl \{  (8 +7e_r^2)(1 -e_r^2)^2
\nonumber \\
& & -{1 \over 336} {\zeta } (1 -e_r^2)
\biggl [ (193\,84 + 470\,4 \eta) + ( 176\,80 +147\,28 \eta)e_r^2
\nonumber \\
& &
- ( 142\,79 - 338\,8\eta)e_r^4 \biggr ]
\nonumber \\
& & +{ \zeta^2 } \biggl [ { 1 \over 181\,44}
\biggl ( ( 381\,349\,6 + 314\,114\,4\eta +725\,76\eta^2)
\nonumber \\
& &
- ( 346\,264\,8  -137\,197\,26 \eta -815\,724\eta^2)e_r^2
\nonumber \\
& & -( 112\,754\,91 - 786\,483\eta -139\,784\,4\eta^2)e_r^4
\nonumber \\
& &
+ ( 357\,872\,4 - 121\,329\,9\eta + 238\,896 \eta^2 )e_r^6
\biggr ) \nonumber \\
& &
-{3 \over 2} ( 1-e_r^2)^{3 \over 2} 
( 5 -2\eta)\,( 8 +7e_r^2)
\biggr ]
\biggr \}.
\end{eqnarray}
	     
We observe that in the test particle limit ( $\eta \rightarrow  0 $)
and  for small radial eccentricities,
Eqs. (\ref{js932}) and (\ref{js929})
become 
\begin{mathletters}
\label{tpl9}
\begin{eqnarray}
<{ d {\cal E} \over d t}>_{\eta=0}&=& {32 \over 5}\,
{c^5\over G}\,{\mu^2 \over m^2}
\,\zeta^5 \biggl \{
1 -{292\,7 \over 336}{\zeta } + {282\,043 \over 907\,2}
{\zeta^2 } 
\nonumber \\
& & + \biggl [ {157 \over 24} - { 639\,7 \over 84} {\zeta }
+ { 273\,523 \over 864} {\zeta^2 } \biggr ]e_r^2
\biggl  \}\,, \\
<{ d {\cal J} \over d t}>_{\eta=0}&=& {32 \over 5}{ \mu^2 \over m}\,c^2
\,\zeta^{7\over 2}
\biggl  \{ 1 - {2423 \over 336}\,{\zeta } 
+ {340\,607 \over 181\,44}\,{\zeta^2 } 
\nonumber \\
& &
+ \biggl [ {23 \over 8} -{ 947\,9 \over 336}\,{\zeta }
+ { 101\,464\,7 \over 181\,44}\,{\zeta^2 } 
\biggr ] e_r^2
\biggr \}\,.
\end{eqnarray}
\end{mathletters}
       Such expressions for average energy and angular momentum
fluxes for a test particle moving in a slightly eccentric
orbit around a Schwarzschild black hole have been obtained by
Tagoshi\cite{HT95}, using the black hole perturbation methods: 
Eqs. (4.9) and (4.12) of~ \cite{HT95}
(with $q =0 $).  They are given by
\begin{mathletters}
\label{htej}
\begin{eqnarray}
<{ d {\cal E} \over d t}> &=&{ 32 \over 5}\,{\mu^2 \over
G\,m^2 \,c^5}\,v^{10}
\biggl \{ 1 -{ 124\,7 \over 336}\,{ v^2 \over c^2}
- {447\,11 \over 907\,2}\,{v^4 \over c^4}
+ \biggl [ { 37 \over 24} -{65 \over 21} {v^2 \over c^2}
- { 474\,409 \over 907\,2}{v^4 \over c^4} \biggr ] e^2
\biggr \}
\,,\\
<{d {\cal J} \over d t}>&=& {32 \over 5}\,{ \mu^2  \over m\,c^5}
\,v^7 \biggl \{ 1 -{ 124\,7 \over 336}\,{ v^2 \over c^2}
- {447\,11 \over 907\,2}\,{v^4 \over c^4}
+ \biggl [ -{ 5\over 8} + { 749 \over 96}\,{ v^2 \over c^2}
-{ 238\,229 \over 604\,8} \,{v^4 \over c^4}
\biggr ]e^2 \biggr \}
\,,
\end{eqnarray}
\end{mathletters}
where $ v$ and $ e$ refer to the radial velocity and
the eccentricity in Schwarzschild coordinates.
Eqs. (\ref{tpl9}) and (\ref{htej})
are consistent, if the ADM variables $a_r$ and $e_r$ are
related to the Schwarzschild variables $v$ and $e$ by 
\begin{mathletters}
\begin{eqnarray}
{ G m \over a_r} &=& v^2\,\biggl  \{ 1 +{v^2 \over c^2} 
+ { 5 \over 4}\,{v^4\over c^4} - \left [ 1 + {v^2 \over c^2}
- { 3329 \over 818}\,{v^4\over c^4} \right ]\,e^2  \biggr \}\,,\\
e_r^2 &=& e^2\,\biggl  \{ 1 + 2\,{v^2 \over c^2} 
+ { 1708 \over 409}\,{v^4\over c^4} \biggr  \}\,.
\end{eqnarray}
\end{mathletters}
As stressed by Tagoshi, the fluxes reveal the more familiar
coefficients in terms of a parameter $v'$, related to the
angular frequency  in the $\phi$ coordinates rather than
$v$, which is adapted to the radial coordinate $r$.
For slightly eccentric orbits, $v$ and $v'$ are related by
\begin{equation}
v = v'\biggl  \{ 1 + {1 \over 2}\left [ 1 -3 {v'^2 \over c^2} 
-12 {v'^4 \over c^4}\right ]e^2 \biggr  \}\,.
\end{equation}
In terms of $v'$ the far-zone fluxes for a test particle in Schwarzschild
geometry, Eqs.( \ref{htej}) 
may be written as
\begin{mathletters}
\label{htfvp}
\begin{eqnarray}
<{ d {\cal E} \over d t}>&=& {32 \over 5}\,{\mu^2 \over G\,m^2 \,c^5}
 v'^{10} \biggl  \{
1 -{124\,7 \over 336}{v'^2 \over c^2} - {447\,11 \over 9072}
{v'^4\over c^4} 
\nonumber \\
& &
+ e^2 \biggl [ { 157 \over 24} -{678\,1 \over 168}{ v'^2 \over c^2}
- { 151\,18 \over 189}{ v'^4 \over c^4}
\biggr ] 
\biggr  \}\,, \\
<{ d {\cal J} \over d t}>&=& {32 \over 5}{ \mu^2 \over m\,c^5}\,v'^7
\biggl \{
1 -{124\,7 \over 336}{v'^2 \over c^2} - {447\,11 \over 9072}
{v'^4\over c^4}
\nonumber \\
& & + e^2\biggl [ { 23 \over 8} -{325\,9 \over 168}{v'^2 \over c^2}
- { 105\, 949\, 3 \over 181\,44}{v'^4 \over c^4}
\biggr ] \biggl \}\,.
\end{eqnarray}
\end{mathletters}
In this form at the Newtonian order, one recovers the 
results of Peters and Mathews \cite{PM63}.  
The quantities $a_r$ and $e_r$ in the ADM 
coordinates  are related to $v'$
and $ e$ by the following relations
\begin{mathletters}
\begin{eqnarray}
{G\,m \over a_r} &=& v'^2 \biggl \{  1 + {1 \over c^2}(1 -2e^2)\,v'^2
+ {1 \over 4\,c^4}\left ( 5 
-{166\,55 \over 409}e^2 \right )\,v'^4 \biggr \} \,,\\
e_r^2 &=& e^2 \biggl \{  1 + 2\,{v'^2 \over c^2}+
{ 1708 \over 409} {v'^4 \over c^4}  \biggr \} \,.
\end{eqnarray}
\end{mathletters}
The above relations may be rewritten, in terms of the conserved 
energy $E$ using \cite{foot}
\begin{mathletters}
\begin{eqnarray}
v^2&=&-2E \biggl \{ 1+e^2 - {E \over 2\,c^2}\biggl (3 - e^2 \biggr )
 +{E^2 \over c^4}\biggl ( 18 +4\,e^2 \biggr ) \biggr \}\,,\\
v'^2&=& -2\,E \biggl \{ 1 -{E \over 2\,c^2}\biggl ( 3 +8\,e^2 \biggr ) 
+{E^2 \over c^4}\biggl (18 +52\,e^2 \biggr ) \biggr \}\,.
\end{eqnarray}
\end{mathletters}
We obtain 
\begin{mathletters}
\begin{eqnarray}
{G\,m \over a_r} &=& v'^2 \biggl \{  1 - {E \over c^2}(2 -4e^2)
+ { E^2 \over c^4}\left ( 8 -{15837 \over 409}e^2 \right ) \biggr \} \,,\\
e_r^2 &=& e^2 \biggl \{  1 -4 {E \over c^2}+
{ 9286  \over 409} {E^2 \over c^4}  \biggr \} \,,
\end{eqnarray}
\end{mathletters}
which are the generalizations  of similar 1PN relations in \cite{JS92}.
\subsection{ The evolution of the orbital elements}
In this section, we compute the 2PN 
corrections to the evolution of orbital elements
due to the emission of gravitational radiation.
We describe  the procedure to compute 
the rate of decrease of the orbital period of the binary in some 
detail and display the final expressions for the 
rate of decay of other elements namely,
$<{ d a_r /dt}>$ and $<{ d e_r /dt}> $.
Employing the heuristic argument, based on the energy and the
angular momentum conservation to the 2PN order, 
the rate of decrease of the orbital
period, $\dot{P}$ of the two compact objects moving,
in quasi-elliptical orbits
is computed. 
The 2PN accurate orbital period, $P=2\pi/n$ given in   
\cite{DS88,SW93,NW95} reads as
\begin{equation}
P = { 2\,\pi\, G\, m \over (-2E)^{3 \over 2}}
\biggl \{ 1 - { 1 \over 4c^2}(15 -\eta)E
-{3 \over 32\,c^4}\biggl [ ( 35 +30\eta +3\eta^2)E^2
-16\,( 5 -2\eta){ (-2E)^{ 3\over 2} \over h }\biggr ] \biggr \}
\label{2pnP}
\end{equation}
Differentiating Eq.~(\ref{2pnP}) with respect to $t$ and
equating  dE/dt to 
$(-<{d{\cal E} / dt} >/ \mu ) $
and ${d h/dt}$ to 
$(-<{d {\cal J}/dt}>/Gm\mu)$ we find
\begin{eqnarray}
\dot P &=& { 6\, \pi\, G\, m \over (-2E)^{5 \over 2}}\biggl \{ 1 -
{ 1 \over 12\, c^2}( 15 - \eta)E + { 1\over 32\, c^4}( 35 +30\eta +3 \eta^2)E^2
\biggr \} < {d {\cal E} \over dt} > 
\nonumber \\
&&-{ 3\,\pi\, \over c^4 \,h^2}(5 -2\eta)
<{d {\cal J} \over dt} >\,.
\label{bs422}
\end{eqnarray}
Note that, in the above equation 
we need $<d{{\cal J}}/dt>$ to the Newtonian accuracy only.
Using in  Eq.~(\ref{bs422}), $<d{\cal E}/dt>$ given by Eq.~(\ref{bs421}) 
and the Newtonian part of Eq.~(\ref{js421}) for $<d{ {\cal J}}/dt>$, we 
get
\begin{eqnarray}
\label{bs426}
\dot P &=& -{192 \over 5}\,{\pi\,\eta }\,
{\zeta^{5 \over 2} \over ( 1-e_r^2)^{7 \over 2}}
\biggl \{  1 + { 73 \over 24}e_r^2 + {37 \over 96}e_r^4
\nonumber \\
& &
- { 1 \over 161\,28}{ \zeta } { 1 \over ( 1-e_r^2)}
\biggl [ ( 598\,56 + 309\,12 \eta) + ( 431\,352 + 134\,848
\eta)e_r^2 
\nonumber \\
& &
+ ( 168\,210 + 556\,08 \eta)e_r^4
- ( 717\,9 - 207\,2\eta )e_r^6 \biggr ]
\nonumber \\
& & + { \zeta ^2 } { 1 \over (1 -e_r^2)^2 }
\biggl [ { 1 \over 580\, 608} \biggl (
(763 \,955\,2 + 607\,737\,6 \,\eta + 483\,840\,\eta^2 )
\nonumber \\
& &
+ ( 263\,832\,80 + 814\,273\,20 \eta + 251\,596\,8 \eta^2 )e_r^2
\nonumber \\
& & - ( 190\,546\,44 - 825\,636\,06 \eta - 170\,553\,6 \eta^2 )e_r^4
\nonumber \\
& &
-( 145\,177\,2  -532\,202\,4\eta - 935\,424 \eta^2 )e_r^6
\nonumber \\
& & + ( 159\,698\,7 - 193\,374 \eta + 745\,92\,\eta^2 )e_r^8
\biggr ) 
\nonumber \\
& &
- { 1 \over 64}\, (5 -2\eta)\,( 1 -e_r^2)^{3 \over 2}
\biggl ( 64 +296\,e_r^2+65 \,e_r^4  
 \biggr )
\biggr ]
\biggr  \}
\end{eqnarray}
Finally inserting  
the expressions for $e_r^2$ and ${ G\,m/a_r}$ in 
terms of E and $h^2$ in Eq. (\ref{bs426})~we obtain  
\begin{eqnarray}
\dot P &=& -{\pi\,\eta \over 5\, c^5}\,{ 1 \over (-E)h^7}
\biggl \{ 425 +732\, Eh^2
+148 E^2h^4 + { 1 \over c^2\,h^2} \left [ { 403\,41 \over 8}
+ { 381\,35 \over 4}\,E h^2 + { 722\,37 \over 14}\,E^2h^4
\right. \nonumber\\
& & \left.
+ { 498\,3\over 7} E^3h^6 -\left ( { 5635 \over 2} + {481\,25 \over 6}\,Eh^2
+ 535\,4\,E^2h^4 + { 140\,6 \over 3}\, E^3h^6\right )\eta \right ]
\nonumber \\
& &
+ { 1 \over c^4}\left [ { 1 \over 672 } \left ( 291\,982\,55
- 309\,096\,90\eta  + 690\,606\,0\eta^2 \right ){1 \over h^4}
\right. \nonumber\\
& & \left.
+ { 1 \over 432}\left ( 293\,418\,53 -505\,570\,59\eta
+ 187\,777\,80 \eta^2 \right ){ E \over  h^2}
+{ 1 \over 2} \left ( 637\,5 -255\,0 \eta \right ) 
{(-2E)^{ 3\over 2} \over h}
\right. \nonumber\\
& & \left.
+ { 1 \over 252}\left ( 864\,965\,0 -219\,467\,70\,\eta 
+ 137\,502\,75\,\eta^2 \right )
\, E^2 
 \right. \nonumber\\
& & \left.
- \left ( 319\,5 -127\,8 \eta \right ) (-2 E)^{5 \over 2}h
+ { 1\over 84}\left ( 166\,451\,5 - 206\,289\,3 \eta + 171\,217\,2 \eta^2
 \right )\, E^3 h^2
 \right. \nonumber\\
&&  \left.
+{ 1 \over 2} \left ( 975 - 390 \eta \right ) \,(-2 E)^{7 \over 2} h^3
+ { 1\over 42} \left ( 163\,085 - 693\,68 \eta + 445\,48 \eta^2 \right )
\, E^4h^4
\right ]
\biggr \} \,.
\end{eqnarray}
In the expression above, $\dot P$ is given as a function 
of the masses and of the 2PN-conserved energy and angular momentum. 
This expression for $\dot{P}$ is
independent of the coordinate system  used to derive it.
Since $P$ is a measurable quantity, one  would have liked
to express $\dot{P}$ in
terms of other directly observable parameters like the orbital period and
some convenient eccentricity  as in the 1PN case\cite{BS89}.
However at present,  to 2PN accuracy  we do not have any such 
suitable and convenient choice
and  therefore we leave the expression for
$\dot{P}$ in terms of the 2PN accurate $E$  and $h^2 $.
	
Similarly, using the definition of $ a_r $ and $ e_r $ 
in terms of E and $ h^2$
and following the method described above, we obtain after a rather
long but straightforward calculation 
\begin{eqnarray}
\label{dadt_2}
<{d a_r \over dt}> &=& -{2 \over 15}\,{ \eta\,  c}
{\zeta ^3 \over (1 -e_r^2)^{11 \over 2}} 
\biggl \{ ( 1 -e_r^2)^2\left ( 96 + 292 e_r^2 + 37 e_r^4 \right  )
\nonumber \\
&& - { 1 \over 56} { \zeta } ( 1 -e_r^2)
\biggl [ ( 280\,16 +940\,8 \eta ) 
+ ( 160\,248 +431\,20\eta )e_r^2 +
\nonumber \\
& &
( 346\,50 + 209\,16 \eta ) e_r^4 
- ( 550\, 1 -103\,6\,\eta )e_r^6 \biggr ]
\nonumber \\
& & + { \zeta ^2 } { 1 \over ( 1-e_r^2)^{11 \over 2}}
\biggl [ 
{ 1\over 604\,8} \biggl ( (137\,748\,16 + 585\,129\,6 \eta +
290\,304 \eta^2 ) 
\nonumber \\
& &
+ ( 428\,878\,40 + 874\,684\,80 \eta + 188\,395\,2 \eta^2)
e_r^2  \nonumber \\
& &
-( 396\,797\,28 - 824\,068\,08\eta -221\,886\,0 \eta^2)e_r^4
\nonumber \\
& &
-( 449\,753\,4 - 103\,086 \eta - 123\,832\,8 \eta^2)e_r^6
\nonumber \\
& &
+ ( 262\,800\,9 - 632\,718 \eta + 839\,16 \eta^2)e_r^8
\biggr )\nonumber \\
& & 
-{ 3 \over 2} ( 1-e_r^2)^{3 \over 2}
\biggl ( ( 5 - 2 \eta )
( 96 + 292\,e_r^2 + 37\,e_r^4 \biggl )
\biggr ]
\biggr \}
\,,\\
<{d e_r \over dt} > &=& -{1 \over 15}\,{c^3 \over G}\,{ \eta \over m}
{ \zeta ^4 \, e_r \over (1 -e_r^2)^{9 \over 2}}
\biggl \{ ( 304 + 121 e_r^2 )(1 -e_r^2)^2
\nonumber \\
& &
- { 1 \over 56} { \zeta }( 1-e_r^2)
\biggl [ ( 133\,640 + 374\,08\eta) +
( 108\,984 + 336\,84 \eta)e_r^2 
\nonumber \\
& &
-( 252\,11 - 338\,8\eta)e_r^4 \biggr ]
\nonumber \\
& &
+ { \zeta^2 } \biggl [
{ 1 \over 201\,6} \biggl ( ( 174\,096\,16 + 170\,583\,84 \eta +
491\,904 \eta^2) \nonumber \\
& &
-( 120\,536\,4 - 397\,143\,72 \eta -760\,788 \eta^2)e_r^2
\nonumber \\
& &
- ( 150\,068\,86 - 224\,584\,2 \eta -560\,952 \eta^2)e_r^4
\nonumber \\
& &
+ ( 384\,043\,5 - 619\,614\eta + 914\,76 \eta^2)e_r^6 \biggr )
\nonumber \\
& & 
- { 3 \over 2} ( 1 -e_r^2)^6\,( 304 + 121e_r^2)\, ( 5 -2\eta)
\biggr ]
\biggr \}
\,.
\end{eqnarray}
To 1PN accuracy we recover the results of \cite{JS92}.
For the special case of circular orbits $ <{d a_r/ dt} >$
takes the simple form  
\begin{equation}
\label{dadtc}
<{d a_r \over dt}> = -{64 \over 5}\,{ \zeta^3 \,\eta \, c}
\biggl \{ 1 -{ \zeta }\left [ {1751 \over 336} + {7
\over 4}\eta \right ]
+ {\zeta^2 }\left [ {294\,383 \over 181\,44}
+ { 263\,65 \over 201\,6}\eta +{ 1 \over 2}\eta^2 \right ]
\biggr \}\,.
\end{equation}
 Eq. (\ref{dadtc}) is consistent with
 the expression for $ \dot r$ given  in
in \cite{GII97}, after taking due account of the coordinate 
transformations required to relate the  ADM and the harmonic gauges
for the circular orbits.
\section{The 2PN contribution to the waveform}
\label{sec:wf2}
In this section, we compute the instantaneous part of 
the 2PN accurate gravitational waveform  {\em i.e.,}  
the transverse - traceless (TT) part of the 2PN accurate far-zone field
for two  compact objects of arbitrary 
mass ratio, moving in a general orbit. 
It is given by \cite{BDI95}:
\begin{eqnarray}
\label{wff}
 (h^{TT}_{km})_{\rm inst} = {2G\over c^4R} {\cal P}_{ijkm}
     \biggl\{ I^{(2)}_{ij}
  &+& {1\over c} \left[ {1\over 3} N_a I^{(3)}_{ija} + {4\over 3}
   \varepsilon_{ab(i} J^{(2)}_{j)a} N_b \right] \nonumber \\
  &+& {1\over c^2} \left[ {1\over 12} N_{ab} I^{(4)}_{ijab} + {1\over 2}
   \varepsilon_{ab(i} J^{(3)}_{j)ac} N_{bc} \right] \nonumber \\
  &+& {1\over c^3} \left[ {1\over 60} N_{abc} I^{(5)}_{ijabc} + {2\over 15}
   \varepsilon_{ab(i} J^{(4)}_{j)acd} N_{bcd} \right] \nonumber \\
  &+& {1\over c^4} \left[ {1\over 360} N_{abcd} I^{(6)}_{ijabcd} + {1\over 36}
   \varepsilon_{ab(i} J^{(5)}_{j)acde} N_{bcde} \right] \biggr\}\,,
\end{eqnarray}
where $ R $ is the Cartesian observer-source distance and 
 $N_a$'s are the  components of ${\bf N }={{\bf X} /R}$,
 the unit normal in the direction
of the  vector $ { \bf X} $, pointing
from the source to the observer.
The transverse traceless 
projection operator projecting orthogonal to  $ { \bf X} $, 
is given by
\begin{equation}
{\cal P}_{ijkm} ({\bf N}) = (\delta_{ik} -N_iN_k)(\delta_{jm}
-N_jN_m)
-{1\over 2} ( \delta_{ij} -N_iN_j) (\delta_{km}-N_kN_m)\ .
\label{eq:2.2}
\end{equation}
Evaluating the appropriate time derivatives of the multipole 
moments  and performing the relevant  
contractions with ${\bf N }$ as required by
Eq. ($\ref{wff} $),
some details of which are given in Appendix A, 
we obtain  
\begin{eqnarray}
\label{giwf1}
\left(h^{TT}_{km}\right)_{inst}
&=& \frac{2G\mu}{c^4R} P_{ijkm}\left\{\xi^{(0)}_{ij} +
{1 \over c}\,\frac{\delta m}{m} \xi^{(0.5)}_{ij} \right. \nonumber \\
&+& {1 \over c^2}\left.\xi^{(1)}_{ij} + 
{ 1 \over c^3}\,\frac{\delta m}{m} \xi^{(1.5)}_{ij} +
{ 1 \over c^4}\,\xi^{(2)}_{ij}\right\}\,,
\end{eqnarray}
where the various $\xi_{ij}$'s are given by
\begin{mathletters}
\label{giwf2}
\begin{eqnarray}
\xi^{(0)}_{ij} &=& 2\biggl( v_{ij} - \frac{Gm}{r}\, n_{ij}\biggr )\,,\\
\xi^{(0.5)}_{ij} &=& \biggl \{ 3 ({\bf N. n}) { G\,m \over r} \left[  
 2n_{(i}v_{j)} -\dot r n_{ij} \right] + 
({\bf N. v})\left [ { G\,m \over r} n_{ij} 
-2 v_{ij} \right ] \biggr \}\,, \\
\xi^{(1)}_{ij} &=&{ 1 \over 3} \biggl \{ (1 -3\eta)
 \biggl [ ({\bf N. n})^2 { G\,m \over r} 
\biggl ( \left ( 3v^2 -15 \dot r^2 + 7 { G\,m \over r} \right )n_{ij} + 30 \dot r n_{(i}v_{j)}
- 14 v_{ij} \biggr )  \nonumber \\
& &+ 
({\bf N. n}) ({\bf N. v}){ G\,m \over r} \left [ 12 \dot r n_{ij}
 -32 n_{(i}v_{j)}
\right ]  
+  ({\bf N. v})^2 \left [ 6 v_{ij} -2{ G\,m\over r} n_{ij} \right ] \biggr ]
\nonumber \\
& & 
+ \biggl [ 3(1 -3\eta) v^2 -2(2 -3\eta){ G\,m \over r} \biggr ] v_{ij}
+ 4 { G\,m \over r} \dot r ( 5 +3\eta) n_{(i}v_{j)} 
\nonumber \\
& &
+ { G\,m \over r} \biggl [ 3( 1-3\eta)\dot r^2 - (10 +3\eta)v^2 
+ 29{ G\,m \over r} \biggr ] n_{ij}
\biggr \} \,,\\
\xi^{(1.5)}_{ij} &=& { 1 \over 12}\,{(1 -2 \eta)}\biggl \{ 
({\bf N. n})^3 { G\,m \over r}
\biggl [   
\left ( 45 v^2 -105 b^2 +90 { G\,m \over r} \right ) \dot r n_{ij}
-96 \dot r v_{ij}
\nonumber \\
& &
-\left ( 42 v^2 
-210 \dot r^2 + 88 { G\,m \over r} \right ) n_{(i}v_{j)} \biggr ]
\nonumber \\
& &
- ({\bf N. n})^2  ({\bf N. v})\,{ G\,m \over r} \biggl [ \left (27 v^2 - 135 \dot r^2 
+84 { G\,m \over r} \right )n_{ij}
+ 336 \dot r n_{(i}v_{j)} -172v_{ij} \biggr ]
\nonumber \\
& &
- ({\bf N. n})({\bf N. v})^2 { G\,m \over r} \biggl [ 48 \dot r n_{ij} -184 n_{(i}v_{j)} \biggr ]
+ ({\bf N. v})^3 \biggl [ 4 { G\,m \over r} n_{ij} -24 v_{ij} \biggr ] \biggl \}
\nonumber \\
& &
- { 1 \over 12}({\bf N. n})\,{ G\,m \over r}
\biggl \{ \biggl [ ( 69 -66 \eta) 
v^2 - (15 -90\eta)\dot r^2 
- ( 242 -24\eta){ G\,m \over r} \biggr ]\dot r n_{ij} 
\nonumber \\
& &
- \biggl [ ( 66 -36\eta)v^2 + (138 + 84\eta)\dot r^2 
\nonumber \\
& &
-( 256 - 72\eta){ G\,m \over r}
\biggr ]n_{(i}v_{j)}
+ ( 192 +12 \eta) \dot r v_{ij} \biggr \}
\nonumber \\
& &
+{ 1 \over 12}\,({\bf N. v})\biggl \{ \biggl [ ( 23 -10\eta)v^2 
-( 9 -18\eta)\dot r^2 
-(104 -12\eta){ G\,m \over r} \biggr ] { G\,m \over r} n_{ij}
\nonumber \\
& &
- \left( 88 +40\eta \right )\,{ G\,m \over r}\,\dot r n_{(i}v_{j)}
- \biggl [ ( 12 -60\eta)v^2 - ( 20 -52\eta) { G\,m \over r} \biggr ]v_{ij}
\biggr \} \,,\\
\xi^{(2)}_{ij} &=& { 1 \over 120}( 1 -5\eta +5 \eta^2) \biggl \{
240\,({ \bf N. v})^4 v_{ij} -({\bf N. n})^4 
\nonumber \\
&&
{ G\,m\over r} \biggl [ \biggl (90 v^4 +
 ( 318 { G\,m\over r} 
-1260 \dot r^2)v^2 
+ 344 { G^2\,m^2 \over r^2} + 1890 \dot r^4 
\nonumber \\
& &
-2310 { G\,m\over r} \dot r^2 \biggr )n_{ij}
\nonumber \\
& &
+ \biggl ( 1620 v^2 +3000 { G\,m\over r} - 3780 \dot r^2 \biggr ) \dot r n_{(i}v_{j)}
- \biggl ( 336 v^2 - 1680 \dot r^2 + 688 { G\,m\over r} \biggr ) v_{ij}
\biggr ]
\nonumber \\
& &
-({\bf N. n})^3 ({ \bf N. v}) { G\,m\over r} \biggl [ \biggl ( 1440 v^2 
- 3360 \dot r^2 +
 3600 { G\,m\over r} \biggr )\dot r n_{ij} 
\nonumber \\
& &
-\biggl ( 1608 v^2 - 8040 \dot r^2 
+ 3864 { G\,m\over r} \biggr ) n_{(i}v_{j)} 
- 3960 \dot r v_{ij} \biggr ]
\nonumber \\
& &
+ 120 ({ \bf N. v})^3 ({\bf N. n}) {G \,m \over r}
 \biggl ( 3 \dot r n_{ij} -20 n_{(i}v_{j)} \biggr )
\nonumber \\
& &
+ ({\bf N. n})^2 ({ \bf N. v})^2 { G\,m\over r} \biggl [ \biggl ( 
396 v^2 -1980 \dot r^2 + 1668 { G\,m\over r} 
\biggr ) n_{ij} + 6480 \dot r n_{(i}v_{j)}
\nonumber \\
& &
-3600 v_{ij} \biggr ] \biggr \}
-{ 1 \over 30}\, ({ \bf N. v})^2 \biggl \{ 
\biggl [ ( 87 -315 \eta +145 \eta^2 )v^2 -( 135 -465\eta + 75 \eta^2 )\dot r^2
\nonumber \\
& &
-( 289 -905\eta +115\eta^2 ){ G\,m\over r} \biggr ] {G\,m\over r}\,n_{ij}
\nonumber \\
& &
- \biggl ( 240 -660\eta -240\eta^2 \biggr )\dot r n_{(i}v_{j)}
\nonumber \\
& &
-\biggl [ ( 30 -270 \eta +630\eta^2)v^2  - 60( 1 - 6\eta 
+10 \eta^2 ){ G\,m\over r}
\biggr ] v_{ij}\biggr \}
\nonumber \\
& &
+ {1 \over 30} ({\bf N. n}) ({ \bf N. v}) { G\,m\over r} 
\biggl \{ \biggl [ ( 270 
- 1140 \eta +1170 \eta^2 )v^2
\nonumber \\
& &
- ( 60 -450 \eta +900 \eta^2 )\dot r^2 -( 1270 -3920 \eta 
+360 \eta^2 ){ G\,m\over r} 
\biggr ] \dot r n_{ij}
\nonumber \\
& &
- \biggl [ ( 186 -810 \eta + 1450\eta^2)v^2  +
 (990 -2910\eta -930\eta^2)\dot r^2 
\nonumber \\
& &
-( 1242 -4170\eta 
+1930\eta^2){ G\,m\over r}
\biggr ]n_{(i}v_{j)}
\nonumber \\
& &
+ \biggl [ 1230 -3810\eta -90\eta^2 \biggr ]\dot r v_{ij} \biggr \}
\nonumber \\
& &
+ { 1 \over 60} ({\bf N. n})^2 { G\,m\over r} \biggl \{ \biggl [
( 117 -480\eta + 540\eta^2 )v^4 -( 630 -2850\eta +4050\eta^2)v^2\dot r^2
\nonumber \\
& &
-( 125 -740\eta +900 \eta^2){ G\,m\over r}\,v^2 
\nonumber \\
& &
+ ( 105 -1050\eta +3150\eta^2)\dot r^4
+( 2715 -8580\eta +1260\eta^2){ G\,m\over r} \,\dot r^2 
\nonumber \\
& &
-( 1048 -3120\eta +240\eta^2) { G^2\,m^2\over r^2} \biggr ] n_{ij}
\nonumber \\
& &
+ \biggl [ ( 216 -1380\eta +4320\eta^2 )v^2 
+ ( 1260 -3300\eta -3600 \eta^2) \dot r^2 
\nonumber \\
& &
-( 3952 - 12860\eta + 3660\eta^2)
{ G\,m\over r} \biggr ]\,\dot r\, n_{(i}v_{j)}
\nonumber \\
& &
- \biggl [ ( 12 -180\eta +1160\eta^2)v^2  
+( 1260 -3840\eta -780\eta^2)\dot r^2 
\nonumber \\
& &
-( 664 -2360\eta +1700\eta^2){ G\,m\over r} \biggr ] v_{ij} \biggr \}
\nonumber \\
& &
- { 1 \over 60}\biggl \{  \biggl [ ( 66 -15 \eta -125\eta^2)v^4
\nonumber \\
& &
+( 90 -180\eta -480\eta^2)v^2 \dot r^2 -( 389 +1030\eta 
-110\eta^2){ G\,m\over r}\,v^2
\nonumber \\
& &
+ ( 45 -225 \eta +225 \eta^2 ) \dot r^4 +
( 915 -1440\eta +720\eta^2){ G\,m\over r}\,\dot r^2 
\nonumber \\
& &
+ ( 1284 +1090\eta){ G^2\,m^2\over r^2} \biggr ]\,{ G\,m\over r}\,n_{ij}
\nonumber \\
& &
-\biggl [ ( 132 +540\eta -580\eta^2 )v^2 
+( 300 - 1140\eta +300\eta^2)\dot r^2
\nonumber \\
& &
+( 856 +400\eta +700\eta^2) { G\,m\over r}
\biggr ]\,{ G\,m\over r }\,\dot r\,n_{(i}v_{j)}
\nonumber \\
& &
-\biggl [ ( 45 -315\eta +585\eta^2 )v^4 
+( 354 -210\eta -550\eta^2)\,{ G\,m\over r}\,v^2 
\nonumber \\
& &
-( 270 -30\eta +270 \eta^2) { G\,m\over r}\,\dot r^2 
\nonumber \\
& &
-( 638 + 1400 \eta -130\eta^2)\,{ G^2\,m^2\over r^2}\biggr ] v_{ij}
\biggr \}\,.
\end{eqnarray}
\end{mathletters}
  The ``tail'' contribution reads 
\begin{eqnarray}
 (h^{TT}_{km})_{\rm tail} = {2G\over c^4R} {2Gm\over c^3}{\cal P}_{ijkm}
\int^{+\infty}_0 &d\tau& \left\{ \ln \left( {\tau\over 2b_1}\right)
    I^{(4)}_{ij} (T_R -\tau)\right. \nonumber\\
 && \quad + {1\over 3c} \ln \left( {\tau\over 2b_2}\right)
  N_a I^{(5)}_{ija} (T_R-\tau) \nonumber\\
 && \quad \left. + {4\over 3c} \ln \left( {\tau\over 2b_3}\right)
 \varepsilon_{ab(i} N_b J^{(4)}_{j)a} (T_R -\tau) \right\} \ ,
 \label{eq:4.3a}
\end{eqnarray}
where we have used for simplicity the notation
\begin{equation}
 b_1 \equiv b\, e^{-11/12}\ , \qquad  b_2 \equiv b\, e^{-97/60}\ , \qquad
 b_3 \equiv b\, e^{-7/6}\ . \label{eq:4.3b}
\end{equation}
We do not discuss the ``tail'' terms in this paper.
Some details of these tail terms may be 
found in \cite{BDI95,WW96}.\\

        The first check on the above waveform is its
circular limit,  which matches with the wave form
computed earlier in \cite{BDI95}.
The next check of the waveform in the general
case is performed by computing the far-zone energy
flux using
\begin{eqnarray}
{d{\cal E}\over dt} = {c^3\,R^2\over 32\pi G} \int \left(
  {\dot h^{TT}_{km}}{\dot h^{TT}_{km}} \right)d\Omega ({\bf N}).
\end{eqnarray}
The expression for $d{\cal E}/dt$ thus obtained is identical to 
the far-zone
energy flux directly obtained from multipole moments Eq.~(\ref{gief}).
Of course, these checks do not uniquely fix the expressions in
Eq.~(\ref{giwf2}) and equivalent expressions are possible leading to the
same transverse traceless parts as discussed below. \\
            
            The above expressions for the waveform,
computed using {\rm STF } multipole
moments differ from 
the corresponding expressions 
 obtained by Will and Wiseman\,
(Eqs. (6.10), (6.11) of \cite{WW96}), using the Epstein-Wagoner 
multipole moments at 1.5PN and 2PN orders.
 Though the two expressions  
are totally different looking at these orders, even in the circular limit, 
it is possible to show that  they are equivalent. 
The equivalence is established by showing that the difference 
between the two expressions, at 1.5PN and 2PN orders has a vanishing 
transverse-traceless,
({\rm TT}) part. The easiest way of  verifying this 
is to show that  the
`plus' and `cross' polarizations of the difference in the two
expressions  vanish at 1.5PN and 2PN orders \cite {WWPC}. In appendix B,
we present the difference -- at 1.5PN and 2PN orders --, 
between our waveform expression
computed directly using the STF multipoles and the Will-Wiseman
one computed using the EW multipoles  and 
verify their equivalence.  
Finally we note that the statement in the appendix
E of \cite {WW96} should more precisely read that, {\rm STF} 
multipole moments presented there
yield an expression for the waveform {\it equivalent} to
Eqs. (6.10) and (6.11) of \cite{WW96},  and not  {\it identical} to it
\cite{WWPC}.

\section{Conclusion}
\label{sec:conc}

                     In this paper using the BDI approach,
we have computed  the 2PN
contributions to the mass quadrupole moment 
for two compact objects of arbitrary mass ratio
moving in general orbits. 
Using this moment we have computed the 2PN 
contributions to the gravitational waveform
and the associated energy and angular momentum fluxes.
These expressions have already proved useful in the
computation of the 2PN radiation reaction, {\em i.e}
the 4.5PN terms in the equations of motion\cite{GII97},
using the refined balance method proposed by Iyer and Will
\cite{IW93,IW95}.
Work is in progress \cite{GI97} to 
obtain  the higher order corrections to the far-zone  
linear momentum flux   
from the gravitational waveform presented here, 
 extending the treatment of Wiseman\cite{AGW93}.
It should be noted that 2PN corrections to the
linear momentum flux can be computed only if one knows
$h^{TT}_{jk} $ to 2.5PN accuracy.  
 Using the 2PN accurate generalized quasi-Keplerian representation
for elliptic orbits, we have computed here the
instantaneous 2PN contributions to $  <{d{\cal E}/dt}>$ 
and $<{d{\cal J}/dt}> $, the fluxes averaged over
one orbital timescale.
This is used to compute the evolution of the orbital elements,
in particular $\dot{P}$, $\dot{e_r}$ and $\dot{a_r}$. 
 The method employed to compute $<{d{\cal E}/dt}>$ and
$<{d{{\cal J}}/dt}>$ could   also be adapted 
to the case of hyperbolic orbits to  generalize the work of 
Simone, Poisson and Will
on the head-on collision \cite{SPW96}. 

\section*{Acknowledgments}
It is a pleasure to thank Luc Blanchet and Thibault Damour
for discussions on the ADM-harmonic transformations.  We
also thank Clifford Will and Alan Wiseman 
for discussions on the waveform equivalence.
\appendix
\section{STF tensors and formulas for the waveform computations}
We present details of the scheme, employed to compute the
contributions to $ h_{jk} $ from various multipole moments,
as required by Eqs. (\ref{wff}),\,(\ref{giwf1})\,and (\ref{giwf2}).
Our scheme proceeds in steps. In the first step, we write
down schematically, the form of the desired
time derivative of the
${\rm STF} $ multipole moment,
using the compact notation $ \{ \} $, introduced by Blanchet
and Damour\cite{bdigen}.
Here $ \{\} $ denotes { \em unnormalised} minimum number of terms,
required  to make the expression symmetric in all the
indicated indices. 
The second step involves {\it peeling}, where by
observation and counting, we rewrite the expression 
obtained in the step 1, as  ${\rm STF} $ on the free indices
-- $ i $ and $ j$ in our case --.
In Step 3, we contract, the final expression of step 2 with
appropriate number of $ {\bf N} $'s as required by Eq. (\ref{wff}).
The actual evaluation of the result of step 3 is
performed on Maple \cite {Maple}.
In all the formulae, $ S_L $, denotes the
symmetric version of the object under consideration; e.g. 
$ S_L=I^{(n)}_{(L)}$ if the object is $ I^{(n)}_L$ and $ S_L= J^{(m)}_{(L)}$
if the object is $J^{(m)}_L$; -- the object in the formula is obvious from
the context.\\
 {\bf The \,\,unnormalized \,\,symmetric \,\,blocks }.
\begin{mathletters}
\begin{eqnarray}
\delta_{\{ij }S_{a\}}&=& \delta_{ij}\,S_{a}
+ \delta_{ia}\,S_{j}  + \delta_{ja}\, S_{i} \,,\\
\delta_{\{ij }S_{ab\}}&=& \delta_{ij}S_{ab} + \delta_{ia}S_{jb}
+ \delta_{ib}S_{aj} +
\nonumber \\
& &
\delta_{ja}S_{ib} + \delta_{jb}S_{ai}
 + \delta_{ab}S_{ij}\,,\\
\delta_{\{ij }\delta_{ab\}}&=& \delta_{ij}\delta_{ab}
+ \delta_{ia}\delta_{jb} +\delta_{ib}\delta_{aj}\,,\\
\delta_{\{ij }S_{abc\}} &=&
\delta_{ij}S_{abc} + \delta_{ia}S_{jbc} +\delta_{ib}S_{ajc}
+\delta_{ic}S_{abj} \nonumber \\
& &
+\delta_{ja}S_{ibc} +\delta_{jb}S_{aic} +\delta_{jc}S_{abi}
\nonumber \\
& &
+\delta_{ab}S_{ijc} +\delta_{ac}S_{ibj} +\delta_{bc}S_{aij} \,,
\\
\delta_{\{ij }\delta_{ab}S_{c \}}&=&
 \biggl \{ \biggl [ \delta_{ja}\delta_{bc} 
+ \delta_{jb}\delta_{ac}
+ \delta_{jc}\delta_{ab}\biggr ]S_{i}
\nonumber  \\
& &+\biggl [ \delta_{ia}\delta_{bc} + \delta_{ib}\delta_{ac}
+ \delta_{ic}\delta_{ab}\biggr ] S_{j}
\nonumber \\
& &+\biggl [ \delta_{ij}\delta_{bc} + \delta_{ib}\delta_{jc}
+ \delta_{ic}\delta_{bj}\biggr ] S_{a}
\nonumber \\
& &+ \biggl [\delta_{ij}\delta_{ac} + \delta_{ia}\delta_{jc}
+ \delta_{ic}\delta_{ja}\biggr ] S_{b}
\nonumber \\
& & + \biggl [\delta_{ij}\delta_{ab} + \delta_{ia}\delta_{jb}
+ \delta_{ib}\delta_{ja}\biggr ] S_{c}\biggr \}\,,\\
\delta_{\{ij }S_{abcd\}} &=&
 \biggl \{ \delta_{ij}S_{abcd} + \delta_{ia}S_{jbcd}
+\delta_{ib}S_{ajcd} +\delta_{ic}S_{abjd}
\nonumber \\
& &
+\delta_{id}S_{abcj} +\delta_{ja}S_{ibcd} +\delta_{jb}S_{aicd}
 +\delta_{jc}S_{abid}
\nonumber \\
& &
+\delta_{jd}S_{abci} +\delta_{ab}S_{ijcd} +\delta_{ac}S_{bdij}
+\delta_{ad}S_{bcij}
\nonumber \\
& &
 +\delta_{bc}S_{adij} +\delta_{bd}S_{acij}
+\delta_{cd}S_{abij} \biggr \}\,,
\\
\delta_{\{ij }\delta_{ab}S_{cd\}} &=&
 \biggl \{ \biggl[ \delta_{ij} \delta_{ab} 
+ \delta_{ia} \delta_{jb}
 +\delta_{ib} \delta_{aj} \biggr ] S_{cd}
+ \biggl[ \delta_{ij} \delta_{ac} + \delta_{ia} \delta_{jc}
 +\delta_{ic} \delta_{aj} \biggr ] S_{bd}
\nonumber \\
& &
+\biggl[ \delta_{ij} \delta_{dc} + \delta_{ic} \delta_{jb}
 +\delta_{ib} \delta_{jc} \biggr ] S_{ad}+
\biggl[ \delta_{ic} \delta_{ab} + \delta_{ia} \delta_{cb}
 +\delta_{ib} \delta_{ac} \biggr ] S_{jd}
\nonumber \\
& &
+\biggl[ \delta_{cj} \delta_{ab} + \delta_{ca} \delta_{jb}
 +\delta_{cb} \delta_{aj} \biggr ] S_{id}
+\biggl[ \delta_{ij} \delta_{ad} + \delta_{ia} \delta_{jd}
 +\delta_{id} \delta_{aj} \biggr ] S_{cb}
\nonumber \\
& &
+\biggl[ \delta_{ij} \delta_{db} + \delta_{id} \delta_{jb}
+\delta_{ib} \delta_{dj} \biggr ] S_{ca}+
\biggl[ \delta_{id} \delta_{ab} + \delta_{ia} \delta_{db}
 +\delta_{ib} \delta_{ad} \biggr ] S_{cj}
\nonumber \\
& &
+\biggl[ \delta_{dj} \delta_{ab} + \delta_{da} \delta_{jb}
 +\delta_{db} \delta_{aj} \biggr ] S_{ci}
+\biggl[ \delta_{ij} \delta_{cd} + \delta_{ic} \delta_{jd}
 +\delta_{id} \delta_{cj} \biggr ] S_{ab}
\nonumber \\
& &
+\biggl[ \delta_{ai} \delta_{cd} + \delta_{ic} \delta_{ad}
 +\delta_{id} \delta_{ac} \biggr ] S_{jb}
+\biggl[ \delta_{aj} \delta_{cd} + \delta_{ca} \delta_{jd}
 +\delta_{ad} \delta_{cj} \biggr ] S_{ib}
\nonumber \\
& &
+\biggl[ \delta_{ib} \delta_{cd} + \delta_{ic} \delta_{db}
 +\delta_{id} \delta_{bc} \biggr ] S_{aj}+
\biggl[ \delta_{bj} \delta_{cd} + \delta_{bd} \delta_{jc}
 +\delta_{jd} \delta_{bc} \biggr ] S_{ai}
\nonumber \\
& &
+\biggl[ \delta_{cd} \delta_{ab} + \delta_{ca} \delta_{db}
 +\delta_{cb} \delta_{ad} \biggr ] S_{ji} \biggr \}\,,
\\
\delta_{\{ij }\delta_{ab}\delta_{cd\}} &=&
 \biggl \{ \biggl [ \delta_{ij}\delta_{ab} 
+ \delta_{ia}\delta_{jb}
+ \delta_{ib}\delta_{ja} \biggr ] \delta_{cd}
+ [ \delta_{ij}\delta_{ac} + \delta_{ia}\delta_{jc}
+ \delta_{ic}\delta_{ja} \biggr ] \delta_{bd}
\nonumber \\
& &
+[ \delta_{ij}\delta_{cb} + \delta_{ic}\delta_{jb}
+ \delta_{ib}\delta_{jc} \biggr ] \delta_{ad}
+ [ \delta_{ic}\delta_{ab} + \delta_{ia}\delta_{cb}
+ \delta_{ib}\delta_{ca} \biggr ] \delta_{jd}
\nonumber \\
& &
+[ \delta_{cj}\delta_{ab} + \delta_{ca}\delta_{jb}
+ \delta_{cb}\delta_{ja} \biggr ] \delta_{id} \biggr \}\,.
\end{eqnarray}
\end{mathletters}
$ {\bf The \,\, {\rm STF} \,\, tensors} $.
\begin{mathletters}
\begin{eqnarray}
{\rm STF}_{ija}( I_{ija}) &=&
S_{ija} - {1 \over 5} \delta_{\{ij }S_{a\}tt}\,,\\
{\rm STF}_{ijab}( I_{ijab}) &=&
S_{ijab} -{ 1 \over 7}\delta_{\{ij }S_{ab\}tt}
+ { 1 \over 35}\delta_{\{ij }\delta_{ab\}}S_{sstt}\,,\\
{\rm STF}_{ijabc}( I_{ijabc}) &=&
S_{ijabc} -{ 1 \over 9}\delta_{\{ij }S_{abc\}tt}
+ { 1 \over 63}\delta_{\{ij }\delta_{ab}S_{c \}sstt}\,,\\
{\rm STF}_{ijabcd}( I_{ijabcd}) &=& S_{ijabcd}
-{ 1 \over 11}\delta_{\{ij }S_{abcd\}pp}
+ { 1 \over 99}\delta_{\{ij }\delta_{ab}S_{cd\}ttqq}
\nonumber \\
& &
-{ 1 \over 693}\delta_{\{ij }\delta_{ab}\delta_{cd\}}S_{ppqqtt}\,.
\end{eqnarray}
\end{mathletters}
{\bf The \,\, `Peeling'}.
\begin{mathletters}
\begin{eqnarray}
{\rm STF}_{ija}( I_{ija}) &=& {\rm STF}_{ij}\biggl \{
S_{ija} -{ 2 \over 5} \delta_{ia} S_{jtt} \biggr \}\,,\\
{\rm STF}_{ijab}( I_{ijab}) &=&
{\rm STF}_{ij}\biggl \{ S_{ijab}
-{ 1\over 7} \biggl [ 2 \delta_{ia}S_{jbtt} +2 \delta_{ib}S_{jatt}
+ \delta_{ba}S_{jitt}\biggr ] +
\nonumber \\
& &
 { 2 \over 35}\biggl [  \delta_{ia}
\delta_{jb} S_{ttss}\biggr ]\biggr \}\,,\\
{\rm STF}_{ijabc}( I_{ijabc}) &=&
{\rm STF}_{ij}\biggl \{ S_{ijabc}
-{ 1\over 9}\biggl [ 2\biggl ( \delta_{ia}S_{jbcpp}
+ \delta_{ib}S_{jacpp} +\delta_{ic}S_{abjpp} \biggr )
\nonumber \\
& &
+ \biggl (\delta_{ab} S_{ijcpp} + \delta_{ac} S_{ijbpp} + \delta_{bc} S_{ijapp}
\biggr ) \biggr ]
\nonumber \\
& &
+ { 2 \over 63} \biggl [  \biggl ( \delta_{ja}\delta_{bc} +
\delta_{jb}\delta_{ac} + \delta_{jc}\delta_{ab} \biggr ) S_{ippqq}
\nonumber \\
& &
+ \biggl ( \delta_{ib}\delta_{jc} S_{appqq} +
\delta_{ia}\delta_{jc} S_{bppqq} + \delta_{ia}\delta_{jb} S_{cppqq}
\biggr ) \biggr ] \biggr \}\,,\\
{\rm STF}_{ijabcd}( I_{ijabcd}) &=&
{\rm STF}_{ij} \biggl \{S_{ijabcd}
-{ 1 \over 11} \biggl [ 2\biggl ( \delta_{ia} S_{jbcdpp}
+ \delta_{ib} S_{jacdpp} + \delta_{ic} S_{abjdpp}
 +\delta_{id} S_{abcjpp} \biggr )
\nonumber \\
& &
+ \biggl ( \delta_{ad} S_{bcijpp} +  \delta_{ab} S_{dcijpp}
\nonumber \\
& &
+\delta_{ac} S_{bdijpp} + \delta_{bc} S_{adijpp}
+ \delta_{bd} S_{acijpp}
+ \delta_{cd} S_{baijpp} \biggr ) \biggr ]
\nonumber \\
& &
+ { 1 \over 99} \biggl [ 2 \biggl (
\delta_{ia}\delta_{jb} S_{cdppqq} + \delta_{ia}\delta_{jc} S_{bdppqq} +
\nonumber \\
& &
\delta_{ic}\delta_{jb} S_{adppqq} + \delta_{ia}\delta_{jd} S_{cbppqq} +
\nonumber \\
& &
\delta_{id}\delta_{jb} S_{acppqq} +
\delta_{ic}\delta_{jd} S_{bappqq}  \biggr )
\nonumber \\
& &
+ 2\biggl ( \delta_{ic} \delta_{ab} + \delta_{ia} \delta_{cb}
+\delta_{ib} \delta_{ca} \biggr ) S_{jdppqq}
+ 2\biggl ( \delta_{id} \delta_{ab} + \delta_{ia} \delta_{db}
+\delta_{ib} \delta_{da} \biggr ) S_{jcppqq}
\nonumber \\
& &
+ 2\biggl ( \delta_{ia} \delta_{cd} + \delta_{ic} \delta_{da}
+\delta_{id} \delta_{ca} \biggr ) S_{jbppqq}
+ 2\biggl ( \delta_{ib} \delta_{cd} + \delta_{ic} \delta_{db}
+\delta_{id} \delta_{bc} \biggr ) S_{jappqq}
\nonumber \\
& &
+ \biggl ( \delta_{cd} \delta_{ab} + \delta_{ca} \delta_{db}
+\delta_{da} \delta_{cb} \biggr ) S_{jippqq} \biggr ]
\nonumber \\
& &
-{ 2 \over 693} \biggl [ \biggl ( \delta_{ia} \delta_{jb} \delta_{cd}
+ \delta_{ia} \delta_{jc} \delta_{bd}
\nonumber \\
& &
+ \delta_{ic} \delta_{jb} \delta_{ad}\biggr )
+ \biggl ( \delta_{ic} \delta_{ab} + \delta_{ia} \delta_{cb} +
\delta_{ib} \delta_{ac} \biggr ) \delta_{jd} \biggr ]
S_{ppqqtt} \biggr \}\,.
\end{eqnarray}
\end{mathletters}
$ {\bf  The \,\, contractions \,\,with }\,\, {\bf N}_{L} $ 
\begin{mathletters}
\begin{eqnarray}
{\rm STF}_{ija}( I^{(3)}_{ija})\,N_a  &=& {\rm STF}_{ij}
\biggl \{ S^{(3)}_{ija}\,N_a -{ 2 \over 5}\,N_i S^{(3)}_{jtt}
\biggr \}\,,\\
{\rm STF}_{ijab}( I^{(4)}_{ijab}) N_{ab} &=&
{\rm STF}_{ij}\biggl \{ S^{(4)}_{ijab}N_{ab}
-{ 1\over 7} \biggl [ 4 N_{ia} S^{(4)}_{jatt} + S^{(4)}_{ijtt} \biggr ]
+ { 2 \over 35} N_{ij} S^{(4)}_{ttss}\biggr \}\,,\\
{\rm STF}_{ijabc}( I^{(5)}_{ijabc})N_{abc} &=&
{\rm STF}_{ij}\biggl \{ S^{(5)}_{ijabc}N_{abc}
-{ 6 \over 9} N_{ibc} S^{(5)}_{jbcpp}
-{ 1 \over 3} S^{(5)}_{ijcpp} N_{c} +
\nonumber \\
& &
 { 6 \over 63} S^{(5)}_{ippqq} N_{j}
+ { 6 \over 63} N_{ija} S^{(5)}_{appqq} \biggr \}\,,\\
{\rm STF}_{ijabcd}( I^{(6)}_{ijabcd})N_{abcd} &=&
{\rm STF}_{ij} \biggl \{ S^{(6)}_{ijabcd}\,N_{abcd}
-{ 8 \over 11} N_{ibcd}\,S^{(6)}_{jbcdpp}
-{6 \over 11} S^{(6)}_{ijcdpp}\, N_{cd}
\nonumber \\
& &
+ { 12 \over 99} N_{ijcd}\, S^{(6)}_{cdppqq} +
{ 24 \over 99} N_{id} \,S^{(6)}_{jdppqq}
+ { 3 \over 99} S^{(6)}_{ijppqq}
\nonumber \\
& &
-{ 12 \over 693} N_{ij}\,S^{(6)}_{ppqqtt} \biggr \}\,.
\end{eqnarray}
\end{mathletters}
$ {\bf The \,\, current\,\, multipole\,\, moments }$.\\
$ \epsilon_{pq(i}\hat {J}_{j)pL}= {\rm STF}_{ij} \biggl \{ \epsilon_{pqi}
\hat {J}_{jpL}\biggr \} $
\begin{mathletters}
\begin{eqnarray}
\epsilon_{pq(i}\hat {J}^{(2)}_{j)p}\,N_q &=&
{\rm STF}_{ij} \biggl \{ \epsilon_{pqi} S^{(2)}_{jp}\,N_q\biggr \}\,,\\
\epsilon_{pq(i}\hat {J}^{(3)}_{j)pa}\,N_{qa}&= &
{ \rm STF}_{ij} \biggl \{ \epsilon_{pqi}
\biggl [ S^{(3)}_{jpa}\,N_{qa}
 -{ 1 \over 5} S^{(3)}_{ptt}\,N_{qj}
\biggr ]\biggr \}\,,\\
\epsilon_{pq(i}\hat {J}^{(4)}_{j)pab}
 \,N_{qab} & =&
{\rm STF}_{ij}\biggl \{ \epsilon_{pqi}\,\biggl [
S^{(4)}_{jpab}\,N_{qab}
-{ 1 \over 7} \biggl ( 2\, S^{(4)}_{pbtt}\,N_{qjb}
+ S^{(4)}_{pjtt}\,N_q \biggr )\biggr ]\biggr \} \,,\\
\epsilon_{pq(i}\hat { J}^{(5)}_{j)pabc}\,N_{qabc} &=&
{\rm STF}_{ij} \biggl \{
\epsilon_{pqi} \biggl [ S^{(5)}_{jpabc}\,N_{qabc}
-{ 1 \over 3} \biggl (  S^{(5)}_{pbctt}\,N_{qjbc}
+  S^{(5)}_{jpctt}\,N_{qc} \biggr )
\nonumber \\
& &
+ { 1 \over 21} \biggl ( S^{(5)}_{pttvv} N_{qj} \biggr )
\biggr ] \biggr \}\,.
\end{eqnarray}
\end{mathletters}
The explicit computations of the above
equations require the
following identities, which are easily derived, using
the rules governing the product of $\epsilon$'s.
The identities are 
\begin{mathletters}
\begin{eqnarray}
{\rm STF}_{ij} \biggl \{ \epsilon_{pqi}\,N_{q}y_j 
{\tilde L_p} \biggr \} &=&
{\rm STF}_{ij} \biggl \{- ({\bf N. v}) y_{ij}+ 
({ \bf N. n})r\,y_{i}v_j \biggr \}
\,,\\
{\rm STF}_{ij} \biggl \{ \epsilon_{pqi}\,N_{q}v_j 
{\tilde L_p} \biggr \}  &=&
{\rm STF}_{ij} \biggl \{ - ({\bf N. v}) y_{i}v_j + 
({ \bf N.n})\,r\,v_{ij} \biggr \}
\,,\\
{\rm STF}_{ij} \biggl \{ \epsilon_{pqi}\,N_{qj}
{\tilde L_p}\biggr \}  &=&
{\rm STF}_{ij} \biggl \{ ({ \bf N.n})\,r\,N_{i}v_{j} -
({\bf N. v})\,y_{i}N_{j} \biggr \}
\,,\\
{\rm STF}_{ij} \biggl \{ \epsilon_{pqi}\,N_{q}y_p 
{\tilde L_j}\biggr \} &=&
{\rm STF}_{ij} \biggl \{ -({\bf N. v})\,y_{ij} + 
({\bf N. n})\,r\, y_iv_j +
( r \dot r ) N_i y_j - r^2 N_iv_j \biggr \} \,,\\
 {\rm STF}_{ij} \biggl \{ \epsilon_{pqi}\,N_{q}v_p 
{\tilde L_j} \biggr \} &=&
{\rm STF}_{ij} \biggl \{ ({ \bf N.n}) \, r\,v_{ij} -({\bf N. v})y_{i}v_j
- r \dot r \,N_{i}v_{j} + v^{2}\, N_{i}y_{j}
\biggr \} \,,\\
{\rm STF}_{ij} \biggl \{ \epsilon_{pqi}\,N_{qj}v_{p}
( {\bf {\tilde L}}.{\bf N}) \biggr \} &=&
{\rm STF}_{ij} \biggl \{ [ v^2 -  ({\bf N. v})^2]N_{i}y_j
+ [ ({\bf N. n})({\bf N. v}) - \dot r] r N_i v_j
\nonumber \\
& &
+ [ \dot r  ({\bf N. v}) - v^2 ({\bf N. n}) ] r N_{ij} \biggr \}\,,\\
{\rm STF}_{ij} \biggl \{ \epsilon_{pqi}\,N_{qj}y_{p}
( {\bf {\tilde L}}.{\bf N})\biggr \} &=&
{\rm STF}_{ij} \biggl \{  [ \dot r -({ \bf N. n }) ({\bf N. v})]r N_{i}y_j
+ [ ({\bf N. n})^2 -1 ] r^2 N_iv_j +
\nonumber \\
& &
[ ({\bf N. v}) - \dot r ({\bf N. n}) ] r^2 N_{ij} \biggr \}\,,\\
{\rm STF}_{ij} \biggl \{ \epsilon_{pqi}\,N_q y_{jp}
( {\bf {\tilde L}}.{\bf N}) \biggr \} &=&
{\rm STF}_{ij} \biggl \{ [ \dot r - ({\bf N. n})  ({\bf N. v}) ]r y_{ij}
+ [ ({\bf N. n})^2 -1 ]r^2 y_iv_j +
\nonumber \\
& &
 [  ({\bf N. v}) - \dot r ({\bf N. n})] r^2 N_iy_j
\biggr \}  \,,\\
{\rm STF}_{ij} \biggl \{ \epsilon_{pqi}\,N_q y_{j}v_{p}
( {\bf {\tilde L}}.{\bf N}) \biggr \}&=&
{\rm STF}_{ij} \biggl \{ [v^2 - ({\bf N. v})^2]y_{ij}
+ [ ({\bf N. n})  ({\bf N. v}) -\dot r]ry_iv_j 
\nonumber \\
& &
+ [ \dot r  ({\bf N. v})
- v^2\,({\bf N. n}) ] r N_iy_j
\biggr \}
\,, \\
{\rm STF}_{ij} \biggl \{ \epsilon_{pqi}\,N_q y_{p}v_j
( {\bf {\tilde L}}.{\bf N}) \biggr \}  &=&
{\rm STF}_{ij} \biggl \{ [\dot r - ({\bf N. n})({\bf N. v})]\,r\,y_{i}v_j
+ [ ({\bf N. n})^2 - 1]\,r^2\,v_{ij}
\nonumber \\
& &
+ [ ({\bf N. v}) - \dot r\,({\bf N. n}) ] r^2\, N_i v_j \biggr \} \,,\\
 {\rm STF}_{ij} \biggl \{ \epsilon_{pqi}\,N_q v_{jp}
( {\bf {\tilde L}}.{\bf N}) \biggr \} &=&
{\rm STF}_{ij} \biggl \{ [ v^2 -  ({\bf N. v})^2]y_{i}v_j
+ [ ({\bf N. n})  ({\bf N. v}) -\dot r ] r v_{ij}
\nonumber \\
& &
+ [ \dot r  ({\bf N. v}) - v^2 ({\bf N. n}) ]r N_iv_j \biggr \}\,,
\end{eqnarray}
\end{mathletters}
where $ {\tilde L_p} =\epsilon_{pkl}\,y_k\,v_l$.

\section{The equivalence to Will-Wiseman waveform}
The expression for the gravitational waveform, 
obtained by Will and Wiseman \cite{WW96}
differs from our waveform expression at the 1.5PN and the 2PN orders.
We give below the difference in the waveform expressions 
at these orders  
and show that the two polarization 
states, $ h_{+} $ and $h_{\times}$ of the difference 
 are zero at 1.5PN and 2PN orders.   
\begin{mathletters}
\begin{eqnarray}
\{(h^{TT}_{km})^{(1.5)}_{BDI} -( h^{TT}_{km})^{(1.5)}_{WW}\} &=& 
{ 1 \over 3\,c^3}\, 
{\cal P}_{ijkm}\, { \delta m \over m}\,
{G\,m \over r} \,
( 1-2\eta) \biggl \{ 
3\, ({ \bf N .n })^3 \dot r v_{ij}
\nonumber \\
& &
-({\bf N . v}) ({ \bf N .n })^2 \biggl [ v_{ij} 
+6 \dot r n_{(i}v_{j)} \biggr ]
\nonumber \\
& &
+ ({ \bf N .n }) ({\bf N . v})^2\,\biggl [ 2\, n_{(i}v_{j)} 
 + 3 \,\dot r\,n_{ij} \biggr ]
\nonumber \\
& &
-({\bf N . v})^3 n_{ij} + 3\,({ \bf N .n }) 
\dot r \biggl [v^2 n_{ij} + v_{ij}
-2\,\dot r\, n_{(i}v_{j)} \biggr ]
\nonumber \\
& &
+({\bf N . v}) \biggl [ v^2 n_{ij} + v_{ij} -2 \dot r n_{(i}v_{j)} 
\biggr ] \biggr \} \,,\\
\{(h^{TT}_{km})^{(2)}_{BDI}-(h^{TT}_{km})^{(2)}_{WW}\} &=& {1 \over 15\,c^4}\,
{\cal P}_{ijkm}
{G\,m \over r} 
\biggl \{ ( 1-5\eta +5\eta^2) 
\biggl [ 12 ({\bf N . v})^4 n_{ij} 
\nonumber \\
& &
- 3 \,({ \bf N .n })^4 \biggl ( 3 v^2 - 
15 \dot r^2 + {G\,m \over r} \biggr )v_{ij}
\nonumber \\
& &
+6 ({ \bf N .n })^3 ({\bf N . v})\biggl ( \left [ 3 v^2 -15\dot r^2 + 
{G\,m \over r} \right ] n_{(i}v_{j)}
-9 \dot r v_{ij} \biggr )
\nonumber \\
& &
- 6 ({ \bf N .n }) ({\bf N . v})^3 \biggl ( 9 \dot r n_{ij} 
+ 4  n_{(i}v_{j)}\biggr )
\nonumber \\
& &
- 3 ({ \bf N .n })^2 ({\bf N . v})^2 \biggl ( \left [ 3 v^2 -15\dot r^2 + {G\,m \over r} \right ]n_{ij}
- 36 \dot r n_{(i}v_{j)} -4 v_{ij} \biggr ]
\nonumber \\
& &
- ({\bf N . v})^2 \biggl [ \biggl ( ( 51 -185 \eta + 55 \eta^2 )v^2 
-( 117 -375 \eta -15\eta^2) \dot r^2 
\nonumber \\
& &
-( 39 -125 \eta -5 \eta^2 )\,{G\,m \over r} \biggr ) n_{ij}
\nonumber \\
& &
-24\,\biggl ( 1 -5\eta +5\eta^2 \biggr ) \dot r n_{(i}v_{j)}
+ 12 \, \biggl ( 1 -5\eta +5\eta^2 \biggr ) v_{ij}
\biggr ] 
\nonumber \\
& &
+ 2 \, ({ \bf N.v})\, ({\bf N. n}) \biggl [ 
27 \, \biggl ( 1 -5\eta +5\eta^2 \biggr )\, \dot r\, v^2 \, n_{ij} 
\nonumber \\
& &
+ \biggl ( ( 39 -125\eta - 5\,\eta^2 ) [ v^2 -  {G\,m \over r} ] 
\nonumber \\
& &
-( 171 -645 \eta + 255\,\eta^2) \dot r^2 \biggr ) n_{(i}v_{j)}
+ 27 \, \biggl ( 1 -5\eta +5\eta^2 \biggr ) \, \dot r \, v_{ij}
\biggr ]
\nonumber \\
& &
- ({\bf N. n})^2 \biggl [ ( 1 -5\eta +5\eta^2 ) 
\biggl ( -9\, v^4 + 45 \dot r^2 v^2 
\nonumber \\
& &
- 3\, v^2\, {G\,m \over r} \biggr ) ( n_{ij} -2\, \dot r\,n_{(i}v_{j)})
\nonumber \\
& &
+ \biggl ( ( 30 -80\eta - 50\,\eta^2 )v^2 
-( 72 -150\eta - 240\,\eta^2 )\dot r^2 
\nonumber \\
& &
-( 42 - 140\eta + 10\,\eta^2){G\,m \over r} \biggr )\,  v_{ij}\biggr ]
\nonumber \\
& &
- \biggl [ \biggl ( (39 - 125\,\eta - 5\,\eta^2 )\,( {G\,m \over r} -\,v^2)
+ ( 117 - 375\eta -15\,\eta^2 )\, \dot r^2 \biggr )
\nonumber \\
& &
\,\biggl ( v^2 \, n_{ij} - 2\,\dot r\,n_{(i}v_{j)} + v_{ij}\biggr )
\biggr ] \biggl \}\,.
\end{eqnarray}
\end{mathletters}
The two independent  
polarization states of the gravitational wave  
$ h_{+} $ and $ h_{\times} $ are given by
$ h_{+} = { 1 \over 2} \biggl ( p_i\,p_j 
-q_i\,q_j \biggr )\,h_{ij}^{TT}$\,
and  
$ h_{\times}  = { 1 \over 2} \biggl ( p_i\,q_j 
+p_j\,q_i \biggr )\,h_{ij}^{TT} $\,, where ${\bf p} $ and ${\bf q}$ 
are the two polarization vectors, forming along with
${\bf N}$ an orthogonal triad \cite{BDI95,BIWW95,WW96}. 
Note that there is no need to apply the {\rm TT } projection
before contracting on ${\bf p} $ and ${\bf q}$.
Consequently, we write the difference in the waveform 
at the 1.5PN and the 2PN orders as
\begin{equation}
\label{nwdfh}
\{(h^{TT}_{ij})_{WW} -( h^{TT}_{ij})_{BDI}\}= \zeta_1 \,v_{ij} + 
\zeta_2\,n_{ij}
+ \zeta_3 \, n_{(i}v_{j)}\,.
\end{equation}
The polarization states $h_{+} $ and $h_{\times}$, for
Eqs. (\ref{nwdfh}) are given by,
\begin{mathletters}
\label{hcpd}
\begin{eqnarray}
h_{+} &=& { 1 \over 2} \biggl ( p_i\,p_j -q_i\,q_j \biggr )\,
\biggl ( \zeta_1 \,v_{ij} + \zeta_2\,n_{ij}
+ \zeta_3 \, n_{(i}v_{j)} \biggr ) 
\,,
\nonumber \\
&= &
{\zeta_1 \over 2} \biggl ( ({\bf p .v })^2 - 
({\bf q.v})^2 \biggr ) + { \zeta_2 \over 2}
\biggl ( ({\bf p.n})^2 - ({\bf q. n})^2 \biggr ) + 
{ \zeta_3 \over 2}\biggl ( ({\bf p.n}) ({\bf p .v}) 
- ({\bf q. n}) ({\bf q.v}) \biggr  )\,,
\\
h_{\times} & =& { 1 \over 2} \biggl ( p_i\,q_j +p_j\,q_i \biggr )\,
\biggl ( \zeta_1 \,v_{ij} + \zeta_2\,n_{ij}
+ \zeta_3 \, n_{(i}v_{j)} \biggr )
\nonumber \\
&=& { \zeta_1 }\,({\bf p.v}) 
\,({\bf q.v}) + \zeta_2 \,({\bf p.n})\,({\bf q. n}) 
+ { \zeta_3 \over 2} \biggl ( ({\bf p.n})\,({\bf q.v}) + 
({\bf p.v})\,({\bf q. n}) \biggr )
\end{eqnarray}
\end{mathletters}
For the explicit computation of Eqs. (\ref{hcpd}), 
we use the standard convention adopted in \cite {BDI95,BIWW95,WW96},
which gives,
${\bf p} = ( 0, 1,0),\,
{\bf q} = ( -\cos i, 0, \sin i),\, 
{\bf N} = ( \sin i, 0, \cos i),\,$
\\
$ {\bf n}= ( \cos \phi, \sin \phi, 0),$ and
${\bf v} = 
( \dot r\,\cos \phi -r\,\omega \,\sin \phi,\, \dot r \,\sin \phi 
 + r\,\omega\,\cos \phi,\, 0) $, \\
where ${\bf n} $ and ${\bf v} $ are  the unit separation vector, 
and the velocity vector respectively,  
$\phi $ is the orbital phase angle, such that the orbital 
angular velocity 
$\omega ={d \phi / dt}$ and 
`$i$' is the inclination
angle of the source. \\
A straightforward but lengthy computation shows that
$h_{+} $ and $ h_{\times}$, given by Eqs. (\ref{hcpd}) vanish, 
both at the  1.5PN and the 2PN orders. This establishes the equivalence of
our waveform expression, Eqs. (\ref{giwf1}) and (\ref{giwf2})
 with the WW one 
given by Eqs. (6.10) and (6.11) of \cite{WW96}.


\begin{references}
\bibitem{foot} Note that  the $E$ denoting conserved energy  in this paper differs from
$E$ in Ref.\cite{HT95} by the rest mass. $E_{Tagoshi}-1=E_{here}$.
\end{references}
\end{document}